\documentclass[aps, pra,twocolumn,superscriptaddress,longbibliography]{revtex4-1}
\usepackage{amsmath}
\usepackage{latexsym}
\usepackage{amssymb}
\usepackage{graphicx,float}
\usepackage{hyperref}
\usepackage{soul}
\usepackage{braket}
\usepackage{verbatim}
\usepackage[italicdiff]{physics}

\hypersetup{
    colorlinks = true,
    linkcolor =blue,
	citecolor=blue, 
	urlcolor=blue 
}

\begin{document}
\title{Quantum non-Markovian Hatano-Nelson model}

\author{Sumit Kumar Jana}
\email{sumitjana05.phy@gmail.com}
\affiliation{Department of Physics, Indian Institute of Technology, Hyderabad 502284, India}

\author{Ryo Hanai}
\email{rhanai09@gmail.com}
\affiliation{Department of Physics, Institute of Science Tokyo, 2-12-1 Ookayama, Meguro-ku, Tokyo, Japan}

\author{Tan Van Vu}
\email{tanvu.tvv@gmail.com}
\affiliation{Center for Gravitational Physics and Quantum Information, Yukawa Institute of Theoretical Physics, Kyoto University, Kitashirakawa-Oiwakecho, Sakyo-ku, Kyoto 606-8502, Japan}

\author{Hisao Hayakawa}
\email{hisao@yukawa.kyoto-u.ac.jp}
\affiliation{Center for Gravitational Physics and Quantum Information, Yukawa Institute of Theoretical Physics, Kyoto University, Kitashirakawa-Oiwakecho, Sakyo-ku, Kyoto 606-8502, Japan}

\author{Archak Purkayastha}
\email{archak.p@phy.iith.ac.in}
\affiliation{Department of Physics, Indian Institute of Technology, Hyderabad 502284, India}

\date{\today} 
\begin{abstract}
{While considering non-Hermitian Hamiltonians arising in the presence of dissipation, in most cases, the dissipation is taken to be frequency independent. 
However, this idealization may not always be applicable in experimental settings, where dissipation can be frequency-dependent. 
Such frequency-dependent dissipation leads to non-Markovian behavior. In this work, we demonstrate how a quantum non-Markovian Hatano-Nelson model arises microscopically in a quasi-one-dimensional dissipative lattice. This is achieved using non-equilibrium Green's functions without requiring any approximation like weak system-bath coupling or a time-scale separation, which would have been necessary for a Markovian treatment. The resulting effective system exhibits nonreciprocal hopping, the defining characteristic of the Hatano-Nelson model, as well as uniform dissipation, both of which are frequency-dependent. This holds for both bosonic and fermionic settings. 
We find solely non-Markovian nonreciprocal features like unidirectional frequency blocking in bosonic settings, and a particular type of non-equilibrium dissipative quantum phase transition in fermionic settings, that cannot be captured in a Markovian theory, nor have any analog in reciprocal systems.
Our results lay the groundwork for describing and engineering non-Markovian nonreciprocal quantum lattices.}
\end{abstract}
\maketitle

{\it Introduction---}The incredibly rich physics of non-Hermitian Hamiltonians, initially proposed as a possible extension of quantum mechanics, has become a direction of mainstream interdisciplinary research, bridging several fields like quantum information, condensed matter, optics, and material science \cite{Ashida_2020, Yang_2023, Zhang_2018}. 
Among these, Hamiltonians with nonreciprocal coupling are especially interesting due to their exotic topological properties, phase transitions, and potential technological advantages in sensing \cite{Zhang_2_2022, Nagaosa_2024, Ding_2022, Wang_2_2023, Zhu_2024, Fruchart_2021, Lau_2018, Bao_2021, Xie_2024, Brighi_2024, McDonald_2020}. 

Systems effectively governed by nonreciprocal Hamiltonians have been realized in several classical platforms like electrical circuits \cite{Liu_2021, Helbig_2020}, metamaterials \cite{Lianchao_2023, Veenstra_2024,Brandenbourger_2019}, photonic crystals \cite{Wang_24}, and acoustic systems \cite{Popa_2014, Shao_2020,rasmussen_2021}. 
These utilize the fact that  Maxwell's equations, Kirchhoff's laws, and Newton's laws in engineered systems can emulate evolution via a non-Hermitian Hamiltonian. 
For quantum nonreciprocal systems, despite a huge body of theoretical work, only a few experimental studies have been possible quite recently \cite{Gou_2020, Xiao_2020, Xiao_2021, Liang_2022, Ren_2023, Wang_2024, Fedorov_2024, Zhao_2025, Shen_2025, Zhang_2025}. 
In the vast majority of theoretical works, nonreciprocal hopping is added phenomenologically by hand to the Hamiltonian. 
However, the realization of quantum nonreciprocal systems requires microscopically deriving the effective non-Hermitian Hamiltonian from standard quantum mechanics. 
Recent works based on Markovian (i.e., having no memory effects) open quantum systems described by the Gorini-Kossakowski-Sudarshan-Lindblad (GKSL) equation \cite{GKSL_1976,Lindblad1976} have established that this is possible in the presence of dissipation and broken time-reversal symmetry \cite{Metelmann_2015, McDonald_2022, Clerk_2022, Wang_2023, Begg_2024, Belyansky_2025, Soares_2025, Brighi_2025}. 
Using this approach, nonreciprocal lattice systems in momentum space have been experimentally realized in ultracold atom platforms \cite{Gou_2020, Liang_2022, Ren_2023, Zhao_2025}. 
Nonreciprocal coupling between a qubit and a resonator \cite{Wang_2024}, as well as in a Josephson ring circuit \cite{Fedorov_2024}, and in a quantum Hall ring \cite{Ochkan_2024} has also been demonstrated. 
 With a related but different approach, the simulation of time evolution via quantum nonreciprocal Hamiltonians has been shown in digital quantum computers \cite{Shen_2025, Zhang_2025} and in photonic quantum walks \cite{Xiao_2020, Xiao_2021}. 

However, barring a few very recent works \cite{luan_2025, Yi_2025, Wilkey_2023_1, Wilkey_2023_2, Lin_2025, Wu_2026, Kuo_2025, Molina_2026}, Markovianity remains a crucial assumption in most microscopic derivations of non-Hermitian Hamiltonians. Markovianity inevitably requires that dissipation be frequency-independent. But this is an idealization, based on approximations like weak system-bath coupling or a time-scale separation between system and bath, which may not always hold in experimental settings \cite{Breuer_2016, Vega_2017, Xu_2010, Liu_2011, Bernardes_2015, Ramos_2016, Cialdi_2019, White_2020, Uriri_2020, Guo_2021, Goswami_2021, Gaikwad_2024, Agarwal_2024, Odeh_2025, Rotter_2009, Longhi_2016, Ordonez_2014, Ordonez_2017, Fang_2017, Huang_2026, Dordevic_2026}. Here, we go beyond this idealization by microscopically designing a nonreciprocal quantum lattice, considering the complete frequency dependence of dissipation. 
This leads to a Hatano-Nelson model \cite{Hatano_1996, Hatano_1997}, having nonreciprocal hopping and uniform dissipation, both frequency dependent, thereby making it non-Markovian.  
Our derivation holds for both bosonic and fermionic systems. 
In the bosonic setting, we find unidirectional frequency blocking, a feature that stems from the interplay of nonreciprocity and frequency-dependent dissipation and is impossible in the Markovian case. In the fermionic setting, we show the existence of a particular type of non-equilibrium dissipative quantum phase transition (NDQPT) 
\cite{Prosen_2008, Prosen_2010,Bastidas_2012, Banchi_2014, Dagvadorj_2015, Carmichael_2015, Archak_2017,Fitzpatrick_2017,Hannukainen_2018, Heugel_2019, Carollo_2020, Zamora_2020, Minganti_2021,Archak_2021, Gamayun_2021, Kuzmin_2025, Saha_2023, Matsumoto_2025}.
This NDQPT has no analog in either Markovian or reciprocal systems.

{\it Hatano-Nelson model and its Markovian derivation---}The Hatano-Nelson model, in open boundary conditions, can be written as
$\hat{H}_{\rm HN}=\sum_{\ell, j=1}^N [\mathbf{H}_{\rm HN}]_{\ell j} \hat{c}_\ell^\dagger \hat{c}_j$, where $[\mathbf{H}_{\rm HN}]_{\ell j}$ is the ($\ell, j$){-}th element of the $N \times N$ single-particle Hamiltonian matrix $\mathbf{H}_{\rm HN}$, given by, 
\begin{align}
	\label{Hatano_Nelson_model}
	[\mathbf{H}_{\rm HN}]_{\ell j}=t_- \delta_{\ell+1,j} + t_+ \delta_{\ell, j+1},~~t_+ \neq t_-^*.
\end{align}
Here, $\hat{c}_j$ ($\hat{c}_j^\dagger$) is a fermionic or bosonic annihilation (creation) operator at the $j$-th site, and $t_+$ ($t_-$) is left-to-right (right-to-left) hopping, whose magnitudes are different, embodying nonreciprocity. When $|t_+| \neq |t_-|$, this nonreciprocity leads to nonreciprocal transport (see Appendix~\ref{Sec:condition for nonreciprocal}), and other remarkable properties like the skin effect \cite{Hatano_1996, Hatano_1997, Zhang_2_2022, Zhu_2024}. Therefore, we focus on the case $|t_+|\ne |t_-|$ in this paper.

It has been recently shown \cite{McDonald_2022, Clerk_2022} that the following GKSL equation can be considered for the effective realization of the Hatano-Nelson model, 
$
\partial_t\hat{\rho} =i[\hat{\rho}, \mathcal{\hat{H}}_S] + \sum_{j=1}^{N-1}( \hat{L}_j \hat{\rho} \hat{L}_j^\dagger - \frac{1}{2}\{ \hat{L}_j^\dagger \hat{L}_j , \hat{\rho} \})$, where 
\begin{align}
\label{Hs}
    \mathcal{\hat{H}}_S=-\sum_{j=1}^{N-1} g(\hat{c}_j^\dagger \hat{c}_{j+1} + \hat{c}_{j+1}^\dagger \hat{c}_j )+\Delta_c \sum_{j=1}^N\hat{c}_j^\dagger \hat{c}_{j},
\end{align}
 is the usual tight-binding Hamiltonian, and $\hat{L}_j=\sqrt{\Gamma}(\hat{c}_j + e^{i\phi}\hat{c}_{j+1})$.  
The Hatano-Nelson model can emerge from this GKSL equation in three different ways: (i) by neglecting the quantum jump terms $\{\hat{L}_j \hat{\rho} \hat{L}_j^\dagger\}$, which corresponds to post-selecting the `no jump trajectory',
(see Appendix~\ref{subsec: HN post selection}), 
(ii) by deriving the evolution equation for the correlation matrix $\langle \hat{c}^\dagger_\ell \hat{c}_j \rangle$, which is governed by an effective non-Hermitian Hamiltonian \cite{McDonald_2022, Archak_2022}; 
(see Appendix~\ref{subsec: HN Lypanunov equation}), 
and (iii) by obtaining the retarded non-equilibrium Green's function (NEGF) in frequency space $\mathbf{G}^{M}(\omega)=[\omega \mathbb{I}-\mathbf{H}_{\rm eff}]^{-1}$, where $\mathbf{H}_{\rm eff}$ is the effective non-Hermitian Hamiltonian and $\mathbb{I}$ is $N \times N$ identity matrix
(see Appendix~\ref{subsec: HN Markovian NEGF}). 
Here and henceforth, the superscript $M$ refers to the Markovian case. 
Each of these yields the same single-particle non-Hermitian Hamiltonian, $\mathbf{H}_{\rm eff}=\mathbf{H}_{\rm HN} +(\Delta_c - i\Gamma) \mathbb{I}+i(\Gamma/2) \mathbf{B}$, with
\begin{align}
	\label{Markovian_non_reciprocal}
	t_{\pm}^{M}=-g-i e^{\mp i\phi} \frac{\Gamma}{2},
\end{align}
and $\mathbf{B}$ contains only boundary terms, $[\mathbf{B}]_{11}=[\mathbf{B}]_{NN}=1$, all other terms of $\mathbf{B}$ being zero. Thus, ignoring the boundary terms, we have the Hatano-Nelson Hamiltonian, along with a decay rate $\Gamma$, which, crucially, does not change the eigenvectors. 
Both the phase $\phi$ and the decay are required for $|t_+| \neq |t_-|$. 
Note that such an effective non-Hermitian Hamiltonian is frequency independent.

\begin{figure}[t!]
	\includegraphics[width=0.9\columnwidth]{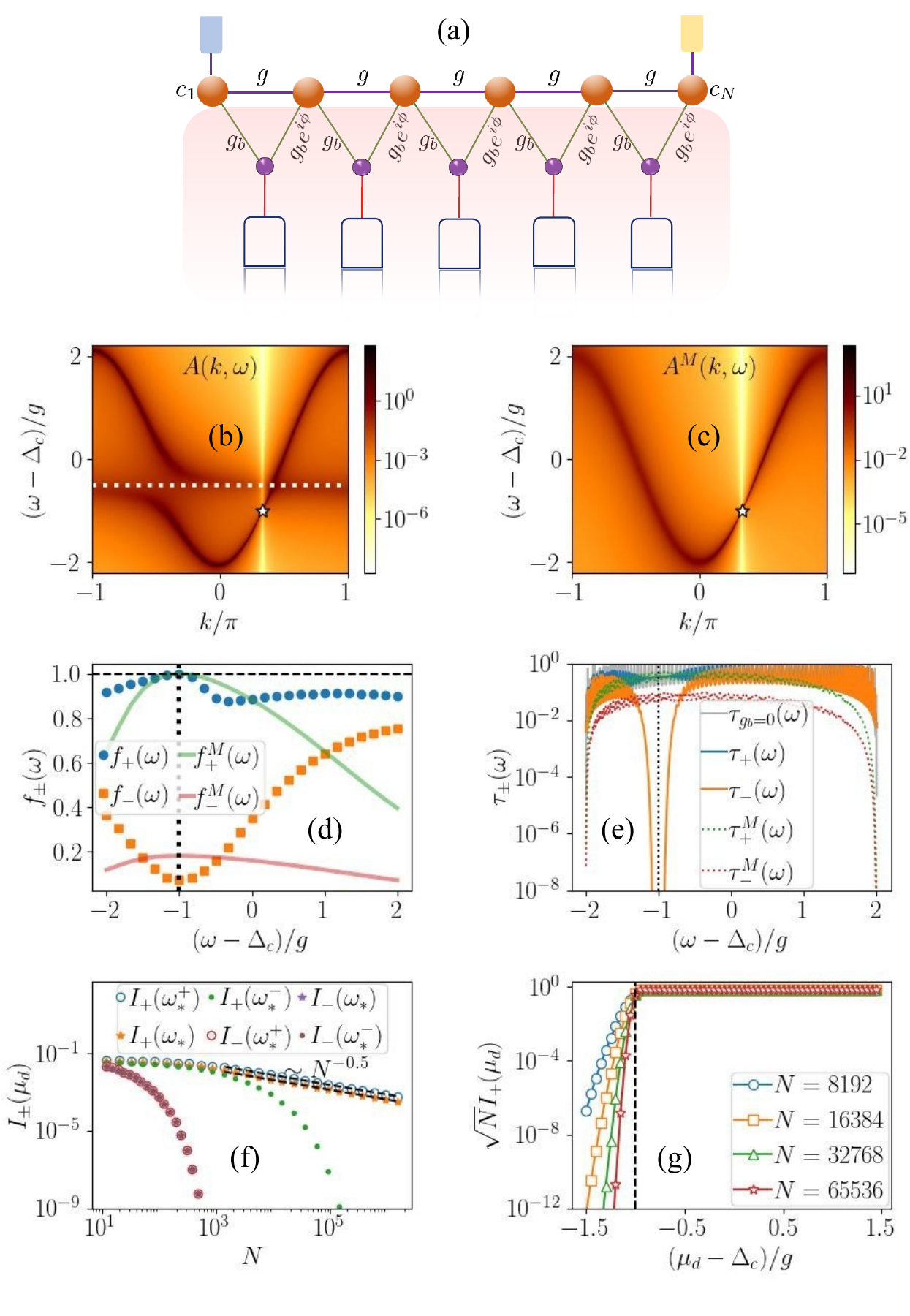}
	\caption{\label{fig:plots} (a) Schematic of our set-up, shown here for $N=6$, see Eqs. \eqref{Hs}, \eqref{setup}. Transport is driven by a coherent drive (bosons) or chemical potential bias (fermions) from the additional leads, shown on top of the chain. Panels (b) and (c) show heatmaps of spectral functions $A(k,\omega)$ and $A^M(k,\omega)$ of non-Markovian and Markovian Hatano-Nelson models, respectively, in the thermodynamic limit. 
		The horizontal dotted line in (b) corresponds to $\omega=\Delta_b$. 
		The stars in (b) and (c) denote the point $(k_*, \omega_*)$. 
		Panel (d) shows plots of $f_{\pm}(\omega)$ (see Eq.~\eqref{G_scaling}), and Markovian counterparts, $f_{\pm}^M(\omega)$. Panel (e) shows transmission amplitude $\tau_+(\omega)$ ($\tau_-(\omega)$) from site $1$ to $N$ ($N$ to $1$), along with their Markovian counterparts, $\tau_{\pm}^M(\omega)$, for $N=100$. The vertical dashed line corresponds to $\omega=\omega_*$.
		The light gray line ($\tau_{g_b=0}(\omega)$) shows the transmission for the reciprocal tight-binding model (i.e, $g_b=0$) for comparison.   
		 Panel (f) shows the scaling of $I_{\pm}(\mu_d)$ with chain length $N$, where $\omega_*^{\pm}=\omega_*\pm 0.1 g$. 
		The black dashed lines are fits to the form $C N^{-0.5}$. We clearly see behavior consistent with Eq.~\eqref{NDQPT}.  Panel (g) shows plots of $\sqrt{N}I_{+}(\mu_d)$ with $\mu_d$. The vertical dashed line shows $\mu_d=\omega_*$. This clearly highlights the NDQPT. 
		Parameters: $\phi=2\pi/3$, $k_*=\pi/3$, $\omega_*=\Delta_c-g$. For all plots except panel (d), $g_b=0.3g$, $\Sigma(z)=\kappa/2$, $\kappa=0.25g$, $\Delta_b=\Delta_c-0.5g$. For panel (e), $g_b=\kappa=0.1g$, $\Delta_b=\omega_*$. In panels (e), (f), (g), $\gamma=0.5g$. In (f), (g), inverse temperature is set to $\beta g=100$. For Markovian plots, $\Gamma(z)=g_b^2/g$.  }
\end{figure}

{\it Non-Markovian microscopic derivation---}We go beyond the Markovian approximation by noting the fact that the Hatano-Nelson model is Gaussian, and NEGF can be derived exactly for Gaussian non-Markovian systems. The key idea is to consider a Gaussian microscopic Hamiltonian of the system and baths, where integrating out the bath degrees of freedom exactly results in the Hatano-Nelson Hamiltonian, up to a uniform on-site term. We find that the set-up schematically described in Fig. \ref{fig:plots}(a) achieves this. It consists of a nearest-neighbor tight-binding chain of $N$ sites. Each pair of neighboring lattice sites is coupled to an auxiliary site, forming a closed triangular unit. Each triangular unit encloses a (synthetic) flux, leading to a complex hopping amplitude in one of the bonds stemming from the Peierls phase factor (see Appendix~\ref{Sec: Hamiltonian for non-Markov}). Each auxiliary site is coupled to its own environment, which gives dissipation. The dissipative auxiliary sites form the engineered bath for the tight-binding chain, which we will integrate out. Such engineered dissipation is, in general, frequency dependent, leading to non-Markovian dynamics. For now, we do not consider the additional leads at the ends of the chain, which are later used to drive transport.

To enable a fully microscopic description, we write down the Hamiltonian for the entire setup, modeling each bath by an infinite number of non-interacting modes. The full Hamiltonian (without the additional left and right leads) is $\hat{\mathcal{H}} = \hat{\mathcal{H}}_S + \sum_{j=1}^{N-1}\left[\hat{\mathcal{H}}_{Sb_j}+ \hat{\mathcal{H}}_{b_j} +\hat{\mathcal{H}}_{bB_j} + \hat{\mathcal{H}}_{B_j}\right]$. Here, the system Hamiltonian $\hat{\mathcal{H}}_S$ is as given in Eq.\eqref{Hs}, and
\begin{align}
	\label{setup}
	&\hat{\mathcal{H}}_{Sb_j}=g_b ( \hat{c}_j^{\dagger} \hat{b}_j + e^{i\phi} \hat{b}_{j}^{\dagger} \hat{c}_{j+1} + {\rm h. c.}  ),~\hat{\mathcal{H}}_{b_j}= \Delta_{b} \hat{b}_j^\dagger \hat{b}_j,  \\
	& 
	\hat{\mathcal{H}}_{bB_j} \!=\! \sum_{r = 1}^{\infty} (\nu_{rj} \hat{b}_j^{\dagger} \hat{B}_{rj} + \nu_{rj}^* \hat{B}_{rj}^{\dagger} \hat{b}_j ),~\hat{\mathcal{H}}_{B_j} = \sum_{r = 1}^{\infty} \Omega_{rj} \hat{B}_{rj}^{\dagger} \hat{B}_{rj} \nonumber, 
\end{align}
where $\hat{b}_j$ is the annihilation operator for $j$-th auxiliary site, $\hat{B}_{rj}$ is the annihilation operator of the $r$-th mode of the bath attached to the $j$-th auxiliary site (see Fig.~\ref{fig:plots}(a)), and h.c. stands for Hermitian conjugate. 
All annihilation operators are either fermionic or bosonic. The relevant properties of the $j$-th bath is entirely governed by its spectral function $\mathcal{J}_j(\omega)=2\pi \sum_{r=1}^\infty |\nu_{rj}|^2 \delta (\omega - \Omega_{rj})$. We consider all baths to have identical spectral functions, $\mathcal{J}_j(\omega)=\mathcal{J}(\omega)$.  At the initial time, we assume no correlations between the auxiliary sites and their respective baths.

In this setting, we integrate out the auxiliary sites along with their baths without any approximation using a quantum Langevin equation approach \cite{Dhar_2006_1, Dhar_2006_2}(see Appendix~\ref{Sec: Non-Markovian derivation}).
This leads to a retarded NEGF in Fourier-Laplace space having the form $\mathbf{G}(z)=[z \mathbb{I}-\mathbf{H}_{\rm eff}(z)]^{-1}$, with the effective non-Hermitian Hamiltonian being  $\mathbf{H}_{\rm eff}(z)= \mathbf{H}_{\rm HN}(z) +[\Delta_c-i \Gamma(z) ] \mathbb{I}+i\left[\Gamma(z)/2\right] \mathbf{B}$, $[\mathbf{H}_{\rm HN}(z)]_{\ell j}= t_-(z) \delta_{\ell+1,j} + t_+(z) \delta_{\ell, j+1}$, where $\mathbf{B}$ is as defined below Eq.~\eqref{Markovian_non_reciprocal}, and
\begin{align}
	\label{NMHN_hoppings}
	t_{\pm}(z) &=-g-ie^{\mp i\phi}\frac{\Gamma(z)}{2},~~\Gamma(z) = \frac{2 g_b^2}{i(\Delta_b - z) + \Sigma(z)},
\end{align}
 $\Sigma(z)=\int_{-\infty}^{\infty} \frac{d\omega}{2\pi} \frac{\mathcal{J}(\omega)}{i(\omega-z)} $ being the self-energy of the baths attached to the auxiliary sites. Comparing with Eq.~\eqref{Markovian_non_reciprocal}, we see that this is similar to the Markovian case, except that $\Gamma(z)$, and hence $t_{\pm}(z)$,  now depend on complex frequency $z$. This frequency dependence, which is a hallmark of non-Markovianity, has important consequences, as we see below.

{\it Thermodynamic limit, dispersion relation and dissipationless nonreciprocal mode---}We first consider the thermodynamic limit, where both left and right edges are taken to infinity. Using translational invariance, we transform to momentum space and obtain the retarded Green's function as $G(k, z) = [z - \varepsilon(k, z)]^{-1}$ with
\begin{align}
	\label{dispersion}
	\varepsilon(k, z)=\Delta_c-2g\cos(k) -i \Gamma(z) \left[1+\cos(k+\phi)\right],	
\end{align}
and the Bloch momentum $k$ is in the Brillouin zone $-\pi \leq k \leq \pi$ (see Appendix~\ref{Sec: Momentum space NEGF}). Here, $\varepsilon(k, z)$ is the effective dispersion relation for the non-Markovian Hatano-Nelson model. It reveals that nonreciprocity microscopically stems from the system experiencing different strengths of dissipation for $k>0$ and $k<0$ when $\phi\neq 0$. Further, we see that at $k=k_*=\pi-\phi$, the dispersion is dissipation-less, leading to a dissipation-less mode $\omega_*=\Delta_c-2g\cos(k_*)$, which is nonreciprocal since $\varepsilon(z,-k_*)$ is not dissipation-less.
Since setting $\Gamma(z)$ to a constant recovers the Markovian case, this mode also exists in the Markovian model. The sign of $k_*$, which is controlled by $\phi$, determines the direction in which transport is preferred. 
For numerical analysis in the following, we fix $\phi=\frac{2\pi}{3}$, which gives $k_*=\frac{\pi}{3}$, $\omega_*=\Delta_c-g$. Thus, transport in the positive $k$ direction (i.e., from left to right) is preferred. In numerical plots, for simplicity, we further set $\Sigma(z)=\kappa/2$ as a constant, although our results remain valid for an arbitrary choice of $\Sigma(z)$.
Note, however, that the decay rate $\Gamma(z)$ (see Eq.~\eqref{NMHN_hoppings})  is still frequency dependent.

To highlight non-Markovian aspects, we plot in Fig.~\ref{fig:plots}(b) the spectral function $A(k, \omega)=-\frac{1}{\pi}{\rm Im}\left[ G(k, \omega) \right]$ for a representative set of parameters, where $\omega$ is real and ${\rm Im}[\ldots]$ denotes the imaginary part. For comparison, the corresponding Markovian spectral function $A^{M}(k,\omega)$ is shown in Fig.~\ref{fig:plots}(c). 
Both Markovian and non-Markovian spectral functions have a deep minimum at $k=k_*$ across all frequencies, except at $\omega=\omega_*$ where there is a high peak.  This shows the dissipationless nature at $\left(k_*,\omega_*\right)$ in both cases. In the non-Markovian case, we additionally observe a clear signature of avoided level crossing at $\omega=\Delta_b$, which is the frequency of the auxiliary site. This occurs because $\Gamma(z)$ exhibits a peak at $\Delta_b$, see Eq.~\eqref{NMHN_hoppings}. Such an avoided level crossing is a hallmark of strong coupling. 
Significant avoided level crossing occurs only for $k<0$, while $k>0$ modes are weakly affected. Thus, close to $\omega=\Delta_b$, $k<0$ modes are strongly coupled to the engineered bath, while $k>0$ modes only experience weak dissipation, resulting in high nonreciprocity. 
Away from this frequency, the system is nearly reciprocal, as shown by approximately equal broadening of peaks for positive and negative momenta. 
By contrast, the Markovian case shows no such features, remaining moderately nonreciprocal at all frequencies, as evidenced by the greater broadening of peaks for $k<0$.

Although the above features are obtained in the thermodynamic limit, they survive and govern the physics in a finite but large enough system with open boundary conditions, as we show below.

{\it Open boundary condition and transfer matrix---}Under open boundary condition, $\mathbf{H}_{\rm eff}(z)$ is tridiagonal and hence $\mathbf{G}(z)$ is the inverse of an $N \times N$ tridiagonal matrix. 
The inverse of a tridiagonal matrix can be easily calculated using the transfer matrix approach \cite{Molinari_1997, Molinari_1998, Lavis_1997, Dwivedi_2016, Kunst_2019, Saha_2023} (see Appendix~\ref{Sec: Transfer Matrix}). 
In our case, the transfer matrix $\mathbf{T}(z)$ is a  $2\times 2$ matrix with elements,  $\left[\mathbf{T}(z)\right]_{11}=z+i\Gamma(z)$, $\left[\mathbf{T}(z)\right]_{12}=-t_+(z)t_-(z)$, $\left[\mathbf{T}(z)\right]_{21}=1$, $\left[\mathbf{T}(z)\right]_{22}=0$. The elements of $\mathbf{G}(z)$ can be obtained in terms of the eigenvalues and eigenvectors of $\mathbf{T}(z)$, which, in a large enough system, leads to the following scaling expressions:
\begin{align}
	\label{G_scaling}
	\Big |[\mathbf{G}(z)]_{\ell j} \Big| \sim 
	\begin{cases}
		& [f_+(z)]^{(\ell-j)}~~{\rm for}~~\ell > j, \\
		& [f_-(z)]^{(j-\ell)}~~{\rm for}~~\ell < j,
	\end{cases}
\end{align}
where $f_{\pm}(z)= |{t_{\pm}(z)}/{\lambda(z)}|$, and $\lambda(z)$ is the eigenvalue of $\mathbf{T}(z)$ with larger magnitude. For real $z=\omega$, $\big | [\mathbf{G}(\omega)]_{\ell j} \big|^2$ is proportional to the transmission probability from site $j$ (input site) to site $\ell$ (output site) for an excitation with frequency $\omega$ (see Appendix~\ref{subsec: bosonic input output}).
Thus, it cannot diverge, and so, from Eq.~\eqref{G_scaling}, $f_{\pm}(\omega) \leq 1$ for real $\omega$. If $f_{\pm}(\omega)  < 1$, the corresponding transmission amplitude decays exponentially with distance between input and output sites. Contrarily, $f_{\pm}(\omega)  = 1$ implies dissipation-less transmission in bulk in the corresponding direction.

In Fig.~\ref{fig:plots}(d), for representative parameters, we show plots of $f_{\pm}(\omega)$, along with their Markovian counterparts, $f_{\pm}^M(\omega)$, obtained by setting $\Gamma(z)$ to constant. We clearly see, $f_{+}(\omega_*)=1$, while $f_{\pm}(\omega)<1$ at all other frequencies. We have verified that varying the parameters does not alter this feature. 
The same holds for the Markovian case, although the overall behavior is different, demonstrating the existence of a dissipation-less nonreciprocal mode in both cases under open boundary conditions. 

Note that the dissipation-less nonreciprocal mode is a property of the nonreciprocal lattice. It does not exist in extremely small systems, for example, in a three- or four-site nonreciprocal system, regardless of non-Markovianity \cite{luan_2025, Yi_2025}. 

{\it Unidirectional frequency blocking in bosonic setting---}The above results directly suggest an interesting application of the bosonic non-Markovian Hatano-Nelson model. 
Consider an input-output experiment with additional probes attached at sites $1$ and $N$ (see Fig.~\ref{fig:plots}(a)). 
A coherent drive at frequency $\omega$ is applied through one probe, and the transmitted radiation is detected at the other (see Appendix~\ref{subsec: bosonic input output}). Our model can be used to strongly attenuate transmission close to a given frequency in one direction, while allowing transmission at all frequencies in the reverse direction. 
Assuming $\Sigma(z)=\kappa/2$ is a constant for simplicity, this occurs when $g\gg g_b \simeq \kappa$, $\Delta_b=\omega_*$. Under this choice, dissipation becomes significant only near $\Delta_b$, making the system strongly nonreciprocal only in the vicinity of width $\sim \kappa$ around this frequency (see Eq.\eqref{NMHN_hoppings}). 
Choosing $\Delta_b=\omega_*$ ensures that, despite nonreciprocity, transmission in the preferred direction remains almost dissipation-less. In Fig.~\ref{fig:plots}(e), we show plots of left-to-right (right-to-left) transmission amplitude $\tau_+(\omega) =\big | [\mathbf{G}(\omega)]_{N 1} \big|^2$ ($\tau_-(\omega)=\big | [\mathbf{G}(\omega)]_{1 N} \big|^2$), along with their Markovian counterparts $\tau_{\pm}^M(\omega)$, and $\tau_{g_b=0}(\omega)$, which is transmission amplitude for reciprocal nearest neighbour tight-binding chain ($g_b=0$ in our setting, see Fig.~\ref{fig:plots}(a)), for representative parameters. 
We clearly observe that near $\omega_*$, $\tau_-(\omega)$ is strongly attenuated, while $\tau_+(\omega)$ is only slightly affected. Away from $\omega_*$, however, the system is nearly reciprocal, since $\tau_-(\omega)$ and $\tau_+(\omega)$ are of similar magnitude, and almost overlap with $\tau_{g_b=0}(\omega)$. In contrast, the Markovian system lacks frequency selectivity and remains moderately nonreciprocal at all frequencies. The oscillations for $|\omega-\Delta_c|< 2g$ seen in the non-Markovian case and the strong attenuation for $|\omega-\Delta_c|> 2g$  seen in all cases are properties of the parent tight-binding model ($g_b=0$) for frequencies within and outside the system band, respectively \cite{Saha_2023}.

{\it NDQPT in fermionic setting---}Taken together, the above results point to an NDQPT in the fermionic setting, with no analog in either the Markovian or reciprocal cases. We consider a fermionic setting at zero temperature and attach two additional baths (see Fig. \ref{fig:plots}(a)): a left lead at chemical potential $\mu_1$ and a right lead at chemical potential $\mu_N$ to drive transport. 
For simplicity, their spectral functions are taken to be constant and equal. 
All bulk baths are assumed to be empty (i.e., with chemical potential $\to -\infty$). 
For left-to-right (right-to-left) transport, we set the chemical potential of the left (right) lead to  $\mu_1 = \mu_d$ ($\mu_N = \mu_d$), while that of the right (left) lead is taken to  $\mu_N \to -\infty$ ($\mu_1 \to -\infty$). 
The resulting particle current into the right (left) lead, denoted by $I_{+}(\mu_d)$ ($I_{-}(\mu_d)$), is evaluated at the nonequilibrium steady state (NESS). 
The NESS currents are expressed in terms of the NEGF (see Appendix~\ref{subsec: fermionic particle current}) as $I_{+}(\mu_d) =\int_{-\infty}^{\mu_d} \frac{d\omega}{2\pi} \gamma^2\Big | [\mathbf{G}(\omega)]_{N 1} \Big|^2 \sim \int_{-\infty}^{\mu_d}d\omega [f_+(\omega)]^{2N} $, $I_{-}(\mu_d) =\int_{-\infty}^{\mu_d} \frac{d\omega}{2\pi} \gamma^2\Big | [\mathbf{G}(\omega)]_{1 N} \Big|^2 \sim \int_{-\infty}^{\mu_d}d\omega [f_-(\omega)]^{2N}$, where $\gamma$ is the constant spectral function of the left and right leads, and we have used Eq.~\eqref{G_scaling}. 
Assuming $k_*>0$, we immediately see that $I_-(\mu_d) \sim e^{-a_-N}$ for $\forall\mu_d$ since $f_-(\omega)<1$ for $\forall\omega$. Contrarily, since $f_+(\omega)$ has a peak at $\omega_*$, we can expand around $\omega_*$ to obtain $I_{+}(\mu_d)\sim \int_{-\infty}^{\mu_d}d\omega e^{N f_+^{\prime \prime}(\omega_*) (\omega - \omega_*)^2}$,  where $f_+^{\prime \prime}(\omega_*)<0$ is the second derivative of $f_+(\omega)$ at $\omega=\omega_*$.
Performing the integral for large enough $N$, we obtain the scaling behavior:
\begin{align}
	\label{NDQPT}
	I_{+}(\mu_d)\sim 
	\begin{cases}
		e^{-a_+N} & \forall\mu_d<\omega_*, \\
		N^{-0.5} & \forall\mu_d\geq\omega_*,
	\end{cases},~ I_{-}(\mu_d)\sim e^{-a_-N}~\forall\mu_d.
\end{align}
Thus, at zero temperature and in the large $N$ limit, a nonanalytic change arises in the behavior of $I_+(\mu_d)$ at $\mu_d=\omega_*$. This is clearly an NDQPT \cite{Archak_2021,Banchi_2014, Carollo_2020, Prosen_2008, Prosen_2010}. But it has no analog in reciprocal systems since  $I_{-}(\mu_d)$ does not show the nonanalytic change.
These features are clearly demonstrated in Figs.~\ref{fig:plots}(f) and \ref{fig:plots}(g).

This NDQPT cannot be captured in the Markovian model.
In the Markovian setting, the only way to attach leads while keeping the effective Hatano-Nelson model undisturbed in the bulk is via the local GKSL approach. As we show in (see Appendix~\ref{Sec: Markovian fermionic current}), in this case we have  $I_{+}^M(\mu_d) \sim N^{-0.5}~\forall\mu_d$ and $I_{-}^M(\mu_d) \sim e^{-a_- N}~\forall\mu_d$, so there is no nonanalytic change in steady state on varying $\mu_d$. 
Thus, the NDQPT is absent in the Markovian model, making it a purely non-Markovian quantum nonreciprocal feature.


We stress that all our analytical results [Eqs.\eqref{NMHN_hoppings} -- \eqref{NDQPT}] are obtained without writing any specific form of the self-energy of the residual bath $\Sigma(z)$ (see Eq.\eqref{NMHN_hoppings}) and therefore hold in complete generality.

{\it Conclusions---}In conclusion, we have obtained quantum non-Markovian Hatano-Nelson model, starting from a fully microscopic Hamiltonian description of the system and baths, without relying on any approximation like weak system-bath coupling or a time-scale separation. The resulting model has nonreciprocal hopping and uniform dissipation, both frequency-dependent. We have revealed solely nonreciprocal non-Markovian NESS features like unidirectional frequency blocking in a bosonic setting and an NDQPT in a fermionic setting.  

By dropping the need to be near the Markovian limit, our results greatly widen the experimental regimes for realizing nonreciprocal quantum lattices, allowing the design and understanding of their non-Markovian versions, a hitherto unexplored direction. Our model may be realized in engineered fermionic or bosonic lattices \cite{Mills_2019, Gou_2020, Borsoi_2024, Yang_2024, Amico_2022, Chien_2015}, while its classical analog may be realized in metamaterials \cite{Lianchao_2023, Veenstra_2024}, photonic crystals \cite{Wang_24}, and acoustic systems \cite{Shao_2020,rasmussen_2021}. 
The unidirectional frequency blocking gives a unique way of controlling the flow of light, while the NDQPT may find applications in quantum sensing \cite{Sarkar_2025, Zhou_2023, Fern_2017, Raghunandan_2018}. Since our description is based on retarded NEGF, it provides a natural language to include many-body interactions via diagrammatic techniques \cite{Jishi_2013, Mattuck_2012}. 
Moreover, the bath initial states do not affect the retarded NEGF. The baths can be at arbitrary temperatures and chemical potentials, or may even be in athermal Gaussian states (example, squeezed states), allowing description of a plethora of non-equilibrium situations. Finally, our fully microscopic understanding enables generalizing to arbitrary non-Markovian nonreciprocal quantum lattices, even in higher dimensions, with potentially rich fundamental physics as well as applications. 


\vspace{1cm}
\textit{Acknowledgments---}SKJ acknowledges financial support from the University Grants Commission(UGC), India (NTA Ref. No. 221610077598). 
SKJ also thanks for the warm hospitality during his stay in the Yukawa Institute for Theoretical Physics, Kyoto University, and the Institute of Science Tokyo.
RH was supported by a Grant in Aid for Transformative Research Areas (No. 25H01364) and for Scientific Research (B) (General) (No. 25K00935). TVV was supported by JSPS KAKENHI Grant Nos.~JP23K13032, JP26K00022, and JP26H02015.
AP thanks the FRIENDSHIP 2.0 program of Japan International Cooperation Agency  (JICA) and IIT Hyderabad Seed Grant for support.  
This work was partially conducted during the long-term workshop ``Frontiers in Non-equilibrium Physics 2024'' (YITP-T-24-01). HH was supported by JSPS KAKENHI Grant No. JP26K06960.

\appendix
\clearpage
\onecolumngrid
\setcounter{equation}{0}




\section*{Appendix}

\section{Condition for nonreciprocal transport}
\label{Sec:condition for nonreciprocal}

 Here we show that $|t_+| = |t_-|$, with $t_+ \neq t_-^*$, i.e, having nonreciprocal hopping only due to a phase factor, does not lead to nonreciprocal transport in the Hatano-Nelson model. In such a system, the eigenvectors are identical to a reciprocal Hamiltonian, while only the eigenvalues acquire a constant phase factor, leading to a lack of nonreciprocal transport features.
 
 We consider a nonreciprocal tight-binding lattice chain of $N$ sites with open boundary conditions, with $t_+=-ge^{i\phi_1}$ and $t_-=-ge^{i\phi_2}$. For $\phi_1 \neq \phi_2$, the system still has nonreciprocal hopping, but, as we will show, it does not have features of nonreciprocal transport. 
\begin{align}
    &\hat{H}_{HN} = -g\sum_{j=1}^{N-1} \left( e^{i\phi_1} \hat{c}_{j}^{\dagger} \hat{c}_{j+1} + e^{i\phi_2} \hat{c}_{j+1}^{\dagger} \hat{c}_{j} \right),
\end{align}
where $\hat{c}_j~(\hat{c}^\dagger_j)$ is the annihilation (creation) operator at $j$-th site.
We can rewrite the Hamiltonian as
\begin{align}
    &\hat{H}_{HN} = e^{i\Phi} \hat{H}', \hat{H}' = -g\sum_{j=1}^{N-1} \left(e^{i\Theta} \hat{c}_{j}^{\dagger} \hat{c}_{j+1} +  e^{-i\Theta} \hat{c}_{j+1}^{\dagger} \hat{c}_{j} \right), ~{\rm where}~ \Phi = \frac{\phi_1+\phi_2}{2},~~\Theta = \frac{\phi_1-\phi_2}{2}.
\end{align}
 In the above, $\hat{H}'$ is a Hermitian Hamiltonian. So, the Hamiltonians $\hat{H}_{HN}$ and $\hat{H}'$ can be diagonalized by the same unitary transformation,
\begin{align}
    \hat{U}^\dagger \hat{H}' \hat{U} = \hat{H}_D,~~\hat{U}^\dagger \hat{H}_{HN} \hat{U} = e^{i\Phi}\hat{H}_D, 
\end{align}
where the columns of the unitary matrix $\hat{U}$ give the eigenvectors of both $\hat{H}'$ and $\hat{H}_{HN}$, and $\hat{H}_D$ is a real diagonal matrix containing the eigenvalues of the Hermitian Hamiltonian $\hat{H}'$. Thus, the eigenvectors of $\hat{H}_{HN}$ do not show any signature of nonreciprocal transport (like skin-effect) under such conditions, since they are the same as those of a Hermitian Hamiltonian. Every eigenvalue gets multiplied by the same complex number, which can at best lead to an overall loss or gain. But, there is no hallmark of nonreciprocal transport.

From above, we conclude that, for nonreciprocal transport in the Hatano-Nelson model, we require
\begin{align}
    |t_+| \neq |t_-|.
\end{align}


\section{Markovian derivation of Hatano-Nelson model}
\label{Sec: Markovian derivation HN}

The Hatano-Nelson model can be shown to arise from the following GKSL equation
\begin{align}
\label{H_NH_Lindblad}
    &\frac{\partial \hat{\rho}}{\partial t}= i[\hat{\rho},\hat{\mathcal{H}}_S]+ \sum_{j} \mathcal{D}^{L}_j[\hat{\rho}(t)],~~\mathcal{D}^{L}_j[\rho(t)] = \hat{L}_j\rho\hat{L}_j^{\dagger} - \frac{1}{2}\{\hat{L}_j^\dagger \hat{L}_j, \hat{\rho} \} \nonumber \\
    &~~\hat{\mathcal{H}}_S = -g\sum_{j=1}^{N-1} ( \hat{c}_{j}^{\dagger} \hat{c}_{j+1} + \hat{c}_{j+1}^{\dagger} \hat{c}_{j} ) +\Delta_c \sum_{j=1}^N\hat{c}_j^\dagger \hat{c}_{j},~~\hat{L}_j=\sqrt{\Gamma}(\hat{c}_j+ e^{i\phi} \hat{c}_{j+1}). 
\end{align}
The Hatano-Nelson model, along with constant dissipation, arises from this GKSL equation in three different ways. 

\subsection{Hatano Nelson model via post-selection}
\label{subsec: HN post selection}

The most standard way to obtain a non-Hermitian Hamiltonian from a GKSL equation is to neglect the jump terms, i.e, terms of the form $\hat{L}_j\rho\hat{L}_j^{\dagger}$. In the language of quantum trajectories, this amounts to post-selecting the no-click trajectory. In our case, this yields
\begin{align}
    &\frac{\partial \hat{\rho}_{\rm ps}}{\partial t}= i\hat{\rho}_{\rm ps}\hat{\mathcal{H}}_{\rm eff}^\dagger-i\hat{\mathcal{H}}_{\rm eff}\hat{\rho}_{\rm ps},~~\hat{\mathcal{H}}_{\rm eff}=\sum_{\ell, m=1}^N \mathbf{H}_{\rm eff}\hat{c_\ell}^\dagger\hat{c}_m,  \\
    & {\rm where}~~\mathbf{H}_{\rm eff}=\mathbf{H}_{\rm HN} +(\Delta_c - i\Gamma) \mathbb{I}+i\frac{\Gamma}{2} \mathbf{B},~~ [\mathbf{H}_{\rm HN}]_{\ell j}=t_- \delta_{\ell+1,j} + t_+ \delta_{\ell, j+1},~~t_{\pm}^M=-g-i e^{\mp i\phi} \frac{\Gamma}{2} \label{H_eff},
\end{align}
$\mathbf{B}$ contains only boundary terms, $[\mathbf{B}]_{11}=[\mathbf{B}]_{NN}=1$, all other terms of $\mathbf{B}$ being zero and $\hat{\rho}_{\rm ps}(t)/{\rm Tr}(\hat{\rho}_{\rm ps}(t))$ is the post-selected density matrix of the no-jump trajectory. The above way of deriving the non-Hermitian Hamiltonian holds for both bosonic and fermionic systems, and is also generalizable to the presence of many-body interactions. But, the probability of observing this trajectory is ${\rm Tr}(\hat{\rho}_{\rm ps}(t)$, which decays exponentially with time, and also with number of sites $N$, making it impractical to realize a nonreciprocal quantum lattice via post-selection. Further, this way of obtaining the non-Hermitian Hamiltonian is crucially related with Markovianity, and is difficult to generalize to non-Markovian systems.

\subsection{Hatano Nelson model via Lyapunov equation}
\label{subsec: HN Lypanunov equation}

The second way to obtain the Hatano-Nelson model in above approach is via obtaining the evolution of the correlation matrx $\mathbf{C}$, whose elements are $[\mathbf{C}]_{\ell m}(t)= {\rm Tr}\left(\hat{c}_\ell^\dagger\hat{c}_m \hat{\rho}(t)\right)$. Since the system is Gaussian, knowing this correlation matrix completely specifies the density matrix. For our case, this evolution equation comes in the form of a Lyapunov equation, 
\begin{align}
\label{HN_Lyapunov}
    \frac{d\mathbf{C}}{dt} = i \mathbf{H}_{\rm eff}^\dagger
    \mathbf{C} - i\mathbf{C}\mathbf{H}_{\rm eff}, 
\end{align}
where $\mathbf{H}_{\rm eff}$ is exactly same as Eq.\eqref{H_eff}, and we have used the fact that in our case, $\mathbf{H}_{\rm eff}^\dagger=\mathbf{H}_{\rm eff}^*$. This does not require post-selection and is scalable. But, it uses the Gaussianity of the system, so would not be applicable if many-body interaction terms were present. Generalizing the Lyapunov equation approach to non-Markovian cases is also not straightforward.

\subsection{Hatano-Nelson model via NEGF}
\label{subsec: HN Markovian NEGF}
The elements of the retarded NEGF in time domain is defined by
\begin{align}
    &\mathcal{G}_{\ell, m}(t)=\langle \{\hat{c}_\ell(t), \hat{c}_m^\dagger(0) \} \rangle~~\textrm{for fermions}, \nonumber \\
    &\mathcal{G}_{\ell, m}(t)=\langle [\hat{c}_\ell(t), \hat{c}_m^\dagger(0) ] \rangle~~\textrm{ for bosons} ,
\end{align}
where the operators are defined in the Heisenberg picture. The GKSL equation is in Schr\"{o}dinger picture. To obtain the retarded NEGF, it is easier to construct a 
 microscopic Hamiltonian model of system coupled to baths, starting from which the same Lypanunov equation, Eq.\eqref{HN_Lyapunov}, can be obtained using a quantum Langevin equation of motion approach. Using the approach in Ref. \cite{Archak_2022} (specifically, using Eqs. (23), (32), (33) of Ref. \cite{Archak_2022}), it can be checked that this leads to the following quantum Langevin equation,
\begin{align}
    \frac{d\hat{c}(t)}{dt}=-i \mathbf{H}_{\rm eff}\hat{c}(t)-i \hat{\eta}(t),
\end{align}
where $\hat{c}(t)$ is a column vector of $N$ elements whose $\ell$th element is $\hat{c}_\ell(t)$, and $ \hat{\eta}(t)$ is a column vector of $N$ elements whose $\ell$th element, $\hat{\eta}_\ell(t)$ is the noise operator due to bath attached at $\ell$th site and satisfies the following properties
\begin{align}
\label{Markovian_QLE_noise}
    \langle\hat{\eta}_m^\dagger(t)\hat{\eta}_\ell(t^\prime)\rangle=0,~~
    \langle \hat{\eta}_\ell(t)\rangle=0,~~ \langle\hat{O}(0)\hat{\eta}_\ell(t)\rangle=\langle\hat{\eta}_\ell(t)\hat{O}(0)\rangle=0,
\end{align}
where $\hat{O}(0)$ is any system operator at initial time. These equations give
\begin{align}
    \frac{d \mathcal{G}(t)}{dt}=-i \mathbf{H}_{\rm eff} \mathcal{G}(t).
\end{align}
Solving this equation via a Fourier-Laplace transform (Laplace transform to variable $s$, followed by $s\to -iz$), leads to the (complex) frequency space retarded NEGF
\begin{align}
    \mathbf{G}^{M}(z)=[z\mathbb{I}- \mathbf{H}_{\rm eff}]^{-1}.
\end{align}
Thus, given the retarded NEGF, $\mathbf{H}_{\rm eff}$ can be identified. Note that all three approaches give the same $\mathbf{H}_{\rm eff}$, and the $\mathbf{H}_{\rm eff}$ is frequency independent in the Markovian case.

Obtaining the non-Hermitian Hamiltonian via NEGF in the above way has the advantage that it can be generalized straightforwardly to non-Markovian situations, since the quantum Langevin equation approach can be used to obtain exact frequency-space retarded Green's functions for any Gaussian system \cite{Dhar_2006_1, Dhar_2006_2}. We obtain the non-Markovian Hatano-Nelson model via this approach.

\section{Hamiltonian for non-Markovian derivation}
\label{Sec: Hamiltonian for non-Markov}

\begin{figure}[H]
    \centering
      \includegraphics[width=1\textwidth]{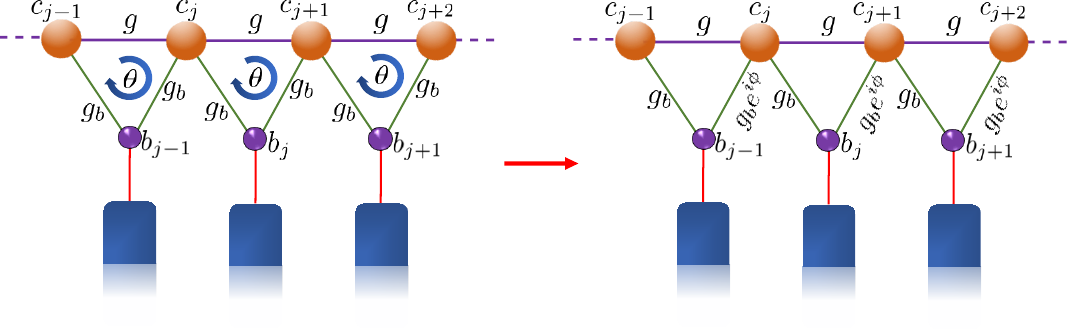}
    \caption{Schematic illustration of the nonreciprocal lattice system. A one-dimensional (1D) lattice chain is considered, where each site coherently interacts with its nearest neighbor with coupling strength $g$. Each pair of neighboring sites simultaneously interacts with a shared local reservoir via a dissipative mode $b_j$. The coupling between the sites and the dissipative mode is described by $g_b$. Using a gauge transformation, the transformed Hamiltonian with a relative phase $\phi=3\theta$ is described on the right side of the schematic.}
    \label{fig. appendix 1}
\end{figure}

In the main text, we have considered the situation given in the right diagram of Fig.\ref{fig. appendix 1}. We arrive at this situation from the situation shown in the left diagram of Fig.\ref{fig. appendix 1}, where there is a flux $\theta$ through each triangular unit. Here we give the details of how this occurs via a gauge transformation.

We consider a nearest-neighbor one-dimensional tight-binding lattice system of $N$ sites. Each pair of neighboring lattice sites is coupled to a common non-Markovian environment, mediated by an auxiliary site, forming a closed triangular unit. Each triangular unit encloses a synthetic Peierls phase factor due to flux, leading to a complex interaction amplitude. However, for illustrative purposes and to deduce the theoretical framework, we consider each reservoir consisting of an infinite number of non-interacting degrees of freedom. The total Hamiltonian of the system-bath composite is structured as 
\begin{align}
 \hat{\mathcal{H}} = \hat{\mathcal{H}}_S + \hat{\mathcal{H}}_{Sb}+ \hat{\mathcal{H}}_b +\hat{\mathcal{H}}_{bB} + \hat{\mathcal{H}}_B .   
\end{align}
Here, the system Hamiltonian $\hat{\mathcal{H}}_S$ is given by
\begin{align}
 \hat{\mathcal{H}}_S = -g\sum_{j=1}^{N-1} ( e^{-i\theta}\hat{c}_{j}^{\dagger} \hat{c}_{j+1} + e^{i\theta}\hat{c}_{j+1}^{\dagger} \hat{c}_{j} ) +\Delta_c \sum_{j=1}^N\hat{c}_j^\dagger \hat{c}_{j} .   
\end{align}
The auxiliary-site Hamiltonian $\hat{\mathcal{H}}_b$ is 
\begin{align}
 \hat{\mathcal{H}}_b=\Delta_{b}\sum_{j=1}^{N-1} \hat{b}_j^\dagger \hat{b}_j  .   
\end{align} 
The coupling between the system and the auxiliary sites is given by
\begin{align}
    \hat{\mathcal{H}}_{Sb}=g_b \sum_{j=1}^{N-1} ( e^{i\theta}\hat{c}_j^{\dagger} \hat{b}_j + e^{i\theta} \hat{b}_{j}^{\dagger} \hat{c}_{j+1} + \text{h.c.}),
\end{align}
while the coupling between the $j$-th auxiliary site and its bath is 
\begin{align}
    \hat{\mathcal{H}}_{bB_j} = \sum_{r = 1}^{\infty} ( \nu_{rj} \hat{b}_j^{\dagger} \hat{B}_{rj} + \nu_{rj}^* \hat{B}_{rj}^{\dagger} \hat{b}_j ).
\end{align}
The Hamiltonian of the corresponding bath is 
\begin{align}
 \hat{\mathcal{H}}_{B_j} = \sum_{r = 1}^{\infty} \Omega_{rj} \hat{B}_{rj}^{\dagger} \hat{B}_{rj},
\end{align}
where $\hat{B}_{rj}$ represents the (fermionic or bosonic) annihilation operator for the $r$-th mode of the bath connected with the $j$-th lattice site.
The Hamiltonians describing the coupling between the auxiliary sites and the baths, as well as the baths themselves, are thus given by
\begin{align}
    \hat{\mathcal{H}}_{bB}&=\sum_{j=1}^{N-1} \hat{\mathcal{H}}_{bB_j}, \quad
    \hat{\mathcal{H}}_B = \sum_{j=1}^{N-1} \hat{\mathcal{H}}_{B_j}.
\end{align}

We apply a gauge transformation to the system bath Hamiltonian to clarify the relation between the main setup and the transformed setup. 
We perform a transformation for both the system operator and also for the auxiliary site operator defined by 
\begin{align}
 \hat{c}_n = e^{in\theta}\hat{\tilde{c}}_n, ~~ \hat{b}_n = e^{i(n-1)\theta}\hat{\tilde{b}}_n.  
\end{align}
This transforms $\hat{\mathcal{H}}_S$ and $\hat{\mathcal{H}}_{Sb}$ to
\begin{align}
    \hat{\mathcal{H}}_S = -g\sum_{j=1}^{N-1} (\hat{c}_{j}^{\dagger} \hat{c}_{j+1} + \hat{c}_{j+1}^{\dagger} \hat{c}_{j} ) +\Delta_c \sum_{j=1}^N\hat{c}_j^\dagger \hat{c}_{j},~~\hat{\mathcal{H}}_{Sb}=g_b \sum_{j=1}^{N-1} ( \hat{c}_j^{\dagger} \hat{b}_j + e^{i\phi} \hat{b}_{j}^{\dagger} \hat{c}_{j+1} + \hat{b}_j^{\dagger} \hat{c}_j + e^{-i\phi} \hat{c}_{j+1}^{\dagger} \hat{b}_{j} ),
\end{align}
with $\phi=3\theta$. This gives the right diagram of Fig.~\ref{fig. appendix 1}, which is used in the main text.

\section{Non-Markovian derivation of Hatano-Nelson model }
\label{Sec: Non-Markovian derivation}

We assume that at initial time, there is no correlation between the baths that give the dissipation to the auxiliary modes and the rest of the set-up.
Without any approximation, quantum Langevin equation approach can be used to obtain the dynamical and the steady-state behavior of the system. The Heisenberg picture can obtain the equation of motion for $\hat{c}_j(t)$. 
\begin{align}
    \label{S9}
    \frac{d\hat{c}_{j}}{dt} = - i\Delta_c \hat{c}_j + ig \hat{c}_{j-1} + ig \hat{c}_{j+1} 
    - ig_b \hat{b}_{j} - ig_b e^{-i\phi} \hat{b}_{j-1} . 
 \end{align}
The quantum Langevin equation can be derived in two steps: 
Initially, one can formally solve the expression for the solution of the annihilation operator of the bath's degree of freedom:
\begin{equation}
    \label{S10}
    \hat{B}_{rj}(t) = e^{-i\Omega_{rj}(t)} \hat{B}_{rj}(0)
    - i\nu_{rj}^* \int_{0}^{t} dt' e^{-i\Omega_{rj} (t-t')} \hat{b}(t')  .
\end{equation}
The bath spectral functions are defined by 
\begin{align}
    \mathcal{J}_{j}(\omega) = 2\pi\sum_{r=1}^{\infty} |\nu_{rj}|^2 \delta(\omega-\Omega_{rj}).
\end{align}
We assume all baths have identical spectral functions,
\begin{align}
    \mathcal{J}_{j}(\omega) =\mathcal{J}(\omega).
\end{align}
Next, substituting this formal solution for the bath operators in the equation of the auxiliary site operator $\hat{b}_j(t)$, the Langevin equation of $\hat{b}_j(t)$ can be derived:
\begin{align}
    \label{S12}
     \frac{d\hat{b}_{j}}{dt} &=-i \Delta_b \hat{b}_j - ig_b\hat{c}_j -ig_b e^{i\phi}\hat{c}_{j+1} - i \sum_{r} \nu_{rj}\hat{B}_{rj} \notag\\
     &= -i \Delta_b \hat{b}_j - ig_b \hat{c}_j -ig_b e^{i\phi}\hat{c}_{j+1} -i\hat{\xi}_j(t) -\int_{0}^{t} dt' \alpha(t-t')  \hat{b}_j(t') ,
\end{align}
where we have introduced
\begin{align}
    &\hat{\xi}_j(t): = \sum_{r=1}^{\infty} \nu_{rj} e^{-i\Omega_{rj} t} \hat{B}_{rj}(0) ,\\
    &\alpha(t-t') := \int \frac{d\omega}{2\pi} \mathcal{J}(\omega) e^{-i\omega(t-t')} 
     .
\end{align}


To obtain the effective part of the system by integrating out the auxiliary modes $\hat{b}_j(t)$. We do so by using Fourier-Laplace transform. We analytically solve Eqs. \eqref{S9} and \eqref{S12} using Laplace transform to variable $s$, substitute the solution of Eq. \eqref{S12} in Eq. \eqref{S9} and eliminate the auxiliary sites . Finally, we put $s=-iz$ ($\hat{o}(s) \rightarrow \hat{o}(z)$) to obtain
\begin{align}
    \label{Main eq}
    &\left( z - \Delta_c + i\Gamma(z) \right) \hat{\tilde{c}}_j(z)+
    \left( g+ ie^{i\phi}\frac{\Gamma(z)}{2} \right) \hat{\tilde{c}}_{j+1}(z)+
    \left( g + ie^{-i\phi}\frac{\Gamma(z)}{2} \right) \hat{\tilde{c}}_{j-1}(z) \notag\\
    &= - ig_b\frac{[\hat{\tilde{\xi}}_j(z) + i\hat{b}_j(0)]}{i(\Delta_b-z) +\Sigma(z)}- ig_be^{-i\phi}\frac{[\hat{\tilde{\xi}}_{j-1}(z) + i\hat{b}_{j-1}(0)]}{i(\Delta_b-z) +\Sigma(z)} + i\hat{c}_j(0), \\
    & \textrm{where } \Gamma(z)=\frac{2g_b^2}{i(\Delta_b-z) +\Sigma(z)}.
\end{align}
where $\hat{\tilde{c}}_j(z)$ is the Fourier-Laplace transform of $\hat{c}_j(t)$, $\hat{\tilde{\xi}}_j(z)$ is the Fourier-Laplace transform of $\hat{\xi}_j(t)$ and  $\Sigma(z)=\int_{-\infty}^{\infty} \frac{d\omega}{2\pi} \frac{\mathcal{J}(\omega)}{i(\omega-z)} $. The above Eq. \eqref{Main eq} can be written as a matrix equation, 
\begin{align}
\label{mat_eq}
    \left[z\mathbb{I}-\mathbf{H}_{\rm eff}(z)\right]\hat{\mathbf{\tilde c}}(z)=\hat{\tilde{\mathbf{\eta}}}(z)
\end{align}
where $\mathbb{I}$ is $N\times N$ identity matrix,  $\hat{\mathbf{\tilde c}}(z)$ is a column vector whose $j$the element is $\hat{\tilde{c}}_j(z)$, $\hat{\tilde{\mathbf{\eta}}}(z)$ is a column vector whose $j$th element is
\begin{align}
    \hat{\eta}_j(z)=- ig_b\frac{[\hat{\tilde{\xi}}_j(z) + i\hat{b}_j(0)]}{i(\Delta_b-z) +\Sigma(z)}- ig_be^{-i\phi}\frac{[\hat{\tilde{\xi}}_{j-1}(z) + i\hat{b}_{j-1}(0)]}{i(\Delta_b-z) +\Sigma(z)} + i\hat{c}_j(0),
\end{align}
and $\mathbf{H}_{\rm eff}(z)$ is the effective non-Hermitian Hamiltonian, given by
\begin{align}
{\mathbf{H}_{\rm eff}(z)} = 
\begin{pmatrix}
\epsilon(z)+i\frac{\Gamma(z)}{2} & t_{-}(z) & 0 & \cdots & 0 \\
t_{+}(z) & \epsilon(z) & t_{-}(z) & \cdots & 0 \\
0 & t_{+}(z) & \epsilon(z) & \cdots & 0 \\
\vdots & \vdots & \vdots & \ddots & \vdots \\
0 & 0 & 0 & \cdots & \epsilon(z)+i\frac{\Gamma(z)}{2}
\end{pmatrix}
\end{align}
where $\epsilon(z) = \Delta_c - i\Gamma(z)$ and  $t_{\pm}(z) = -g- ie^{\mp i\phi}\frac{\Gamma(z)}{2}$. From Eq.\eqref{mat_eq}, solving for $\hat{\mathbf{\tilde c}}$ gives
\begin{align}
    \hat{\mathbf{\tilde c}}(z)=\mathbf{G}(z)\hat{\mathbf{\eta}}(z),
\end{align}
where 
\begin{align}
    \mathbf{G}(z)=\left[z\mathbb{I}-\mathbf{H}_{\rm eff}(z)\right]^{-1},
\end{align}
is the retarded NEGF of the system. The frequency dependence of $\mathbf{H}_{\rm eff}(z)$ embodies the non-Markovian nature of the dynamics.

\section{Momentum space NEGF in thermodynamic limit}
\label{Sec: Momentum space NEGF}

The above derivation gives the NEGF in open boundary condition of a finite system. In the thermodynamic limit, we can use translational invariance to obtain the NEGF in momentum-frequency space. The transformation to momentum space is given by $d_j(t)=\int_{-\infty}^{\infty} \frac{dk}{2\pi} e^{ikj}\hat{d}(k, t)$, where $\hat{d}_j$ is the annihilation operator for either the system sites or the auxiliary sites. Using this transformation on Eq. \eqref{Main eq} gives 
\begin{align}
    [z - \Delta_c + 2g\cos(k) + i\Gamma(z)(1+\cos(k+\phi))]  \hat{\tilde{c}}(k,z)  = -ig_b\frac{\hat{\tilde{\xi}}(k,z)+i\hat{b}(k,0)}{i(\Delta_b-z) +\Sigma(z)}[1+e^{-i(k+\phi)}] + i\hat{c}(k,0),
\end{align}
We obtain the retarded NEGF in momentum-frequency space as $G(k, z) = [z - \varepsilon(k, z)]^{-1}$ with
\begin{align}
	\label{dis}
	\varepsilon(k, z)=\Delta_c-2g\cos(k) -i \Gamma(z) \left[1+\cos(k+\phi)\right].
\end{align}

\section{The transfer matrix and the NEGF scaling relations}
\label{Sec: Transfer Matrix}

The elements of the NEGF under open boundary conditions can be obtained via a transfer matrix approach. 
$\mathbf{G}(z)$ is an inverse of a tri-diagonal matrix $\mathcal{M}(z)$. Using result for inverse of tridiagonal matrix, the element of $\mathbf{G}(z)$ are given by
\begin{align}
    \label{G elems}
	[\mathbf{G}(z)]_{\ell j} = 
	\begin{cases}    
		& (-1)^{l+j} \frac{(-t_+)^{\ell-j}\tilde{\Delta}_{j-1}(z) \Delta_{\ell+1}(z)}{\tilde{\Delta}_N(z)}~~{\rm for}~~\ell > j\\
        & \frac{{\tilde{\Delta}}_{\ell-1}\Delta_{j+1}}{{\tilde{\Delta}_N(z)}}~~{\rm for}~~\ell=j \\
		& (-1)^{l+j} \frac{(-t_-)^{j-\ell}\tilde{\Delta}_{\ell-1}(z) \Delta_{j+1}(z)}{\tilde{\Delta}_N(z)}~~{\rm for}~~\ell < j
	\end{cases} 
    .
\end{align}
where, 
\begin{align}
\label{Delta}
    &\Delta_1(z) = \left[z - \epsilon(z) -\Sigma_{11}(z) \right]\Delta_2(z) - t_{-}(z)t_{+}(z)\Delta_3(z) \notag ,\\
    &\Delta_N(z)=z-\epsilon(z)-\Sigma_{NN}(z), \\
    &\Delta_{N+1}(z)=1, \nonumber \\
    &  \begin{pmatrix}
        \Delta_i(z)\\ \Delta_{i+1}(z)
    \end{pmatrix} = \mathbf{T}(z)
    \begin{pmatrix} \Delta_{i+1}(z) \notag\\
    \Delta_{i+2}(z) 
    \end{pmatrix}
    ~~{\rm for }~~2 \leq i \leq N - 1,
\end{align}
and 
\begin{align}
\label{Delta_tilde}
    & \tilde{\Delta}_1(z) = z - \epsilon(z)-\Sigma_{N N}(z), \nonumber\\
    & \tilde{\Delta}_{0}(z)= 1 \nonumber\\
    & \tilde{\Delta}_{N}(z)=\left[z - \epsilon(z) -\Sigma_{11}(z) \right]\tilde{\Delta}_{N-1}(z) - t_{-}(z)t_{+}(z)\tilde{\Delta}_{N-2}(z) \\
    &  \begin{pmatrix}
        \tilde{\Delta}_i(z)\\ \tilde{\Delta}_{i-1}(z)
    \end{pmatrix} = \mathbf{T}(z)
    \begin{pmatrix}\tilde{\Delta}_{i-1}(z) \notag\\
    \tilde{\Delta}_{i-2}(z) 
    \end{pmatrix}
     ~~{\rm for}~~ 2\leq i\leq N-1, \nonumber
\end{align}
with 
\begin{equation}
\label{TM}
    \mathbf{T}(z) = \begin{pmatrix}
    z - \epsilon(z) & -t_{-}(z)t_{+}(z) \\
    1 & 0
\end{pmatrix}
\end{equation}
is the transfer matrix. Here, we have absorbed the boundary terms into $\Sigma_{11}(z)$, $\Sigma_{NN}(z)$, which may also include contributions from additional leads (see later). Diagonalizing the transfer matrix and using the recursion relations in Eqs.\eqref{Delta} and \eqref{Delta_tilde}, it can be checked that, for $\ell$ away from the boundaries, we have the scaling relations
\begin{align}
    \tilde{\Delta}_\ell(z) \sim [\lambda(z)]^{\ell},~~\Delta_\ell(z) \sim [\lambda(z)]^{N-\ell},
\end{align}
where, $\lambda(z)$ is the eigenvalue of $\mathbf{T}(z)$ with higher magnitude. Using these in Eq.\eqref{G elems}, we get the scaling relations
\begin{align}
    \Big |[\mathbf{G}(z)]_{\ell j} \Big| \sim 
	\begin{cases}
		& [f_+(z)]^{(\ell-j)}~~{\rm for}~~\ell > j, \\
		& [f_-(z)]^{(j-\ell)}~~{\rm for}~~\ell < j,
	\end{cases}, ~~{\rm where}~~f_{\pm}(z)= \left|\frac{t_{\pm}(z)}{\lambda(z)}\right|,
\end{align}
which are used in the main text.

The Markovian case simply corresponds to $t_{\pm}(z) \to t_{\pm}^M$, independent of $z$. So, for Markovian case
\begin{align}
    \Big |[\mathbf{G}^M(z)]_{\ell j} \Big| \sim 
	\begin{cases}
		& [f_+^M(z)]^{(\ell-j)}~~{\rm for}~~\ell > j, \\
		& [f_-^M(z)]^{(j-\ell)}~~{\rm for}~~\ell < j,
	\end{cases}, ~~{\rm where}~~f_{\pm}(z)= \left|\frac{t_{\pm}^M}{\lambda(z)}\right|.
\end{align}

\section{Driving transport via two additional leads}
\label{Sec: Two additional leads}

\begin{figure}[H]
    \centering
      \includegraphics[width=0.6\textwidth]{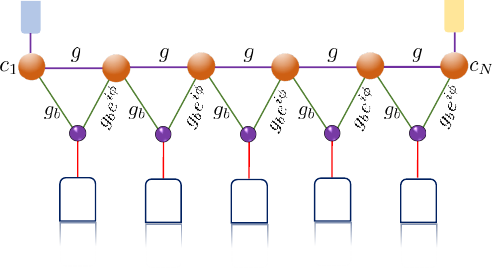}
    \caption{We connect our set-up to two more baths, one at site $1$ and another at site $N$ to drive transport, in both bosonic and fermionic settings.}
    \label{fig. two_additional_leads}
\end{figure}

To obtain the NESS transport behavior, we connect two additional leads at its two sites $1$ and $N$, see Fig.~\ref{fig. two_additional_leads}. The full system bath Hamiltonian is given by
\begin{align}
 & \hat{\mathcal{H}} = \hat{\mathcal{H}}_S + \hat{\mathcal{H}}_{Sb}+ \hat{\mathcal{H}}_b +\hat{\mathcal{H}}_{bB} + \hat{\mathcal{H}}_B + \sum_{j=1,N}\hat{H}_{SB_j}+\sum_{j=1,N}\hat{H}_{B_j}, \\
 & \hat{H}_{SB_j} = \sum_{r = 1}^{\infty} ( \gamma_{rj} \hat{c}_j^{\dagger} \hat{\mathcal{B}}_{rj} + \gamma_{rj}^* \hat{\mathcal{B}}_{rj}^{\dagger} \hat{c}_j ),~~\hat{H}_{B_j} = \sum_{r = 1}^{\infty} \omega_{rj} \hat{\mathcal{B}}_{rj}^{\dagger} \hat{\mathcal{B}}_{rj}.
\end{align}
Here $\hat{H}_{SB_j}$ is the coupling Hamiltonian for interaction with boundary leads and $\hat{H}_{B_j}$ is the Hamiltonian of the corresponding lead. This modifies the Green's function as $\mathbf{G}(z) = [\mathcal{M}(z)]^{-1}$, where $\mathcal{M}(z)= z \mathbb{I}-\mathbf{H}_{\rm eff}(z)-\mathbf\Sigma(z)$. The form of $\mathcal{M }(z)$ is given by
\begin{align}
\label{modified G}
{\mathcal{M}(z)} = 
\begin{pmatrix}
z-\epsilon(z) -\Sigma_{11}(z) & -t_{-}(z)& 0 & 0 & \cdots \\
-t_{+}(z) & z-\epsilon(z)& -t_{-}(z) &0 & \cdots \\
0 & -t_{+}(z) & z-\epsilon(z)& -t_{-}(z)& \cdots  \\
\vdots & \vdots & \vdots& \vdots&\vdots \\
0 & 0& 0&z-\epsilon(z)&-t_{-}(z)\\
0 & 0 & 0& -t_{+}(z)& z-\epsilon(z) -\Sigma_{NN}(z)
\end{pmatrix}.
\end{align}
Here, $\Sigma_{11}(z)$ and $\Sigma_{NN}(z)$ describe both, the self-energy term due to the presence of the two leads and the contribution from boundary terms. For simplicity, we use the wide-band limit, these terms simplify to 
\begin{align}
 \mathbf\Sigma_{11}(z) = -i\gamma/2+ i\frac{\Gamma(z)}{2},~~ \mathbf\Sigma_{NN}(z) = -i\gamma/2+ i\frac{\Gamma(z)}{2}.   
\end{align}
Additionally, the first and $N$th terms of the vector $\hat{\mathbf{\eta}}(z)$ in Eq.\eqref{mat_eq} changes as
\begin{align}
     \hat{\eta}_j(z)\to \hat{\eta}_j(z)+\hat{\tilde{\zeta}}_j(z),~~{\rm for}~~j=1,N,
\end{align}
where the noise due to the probes is
\begin{align}
    \hat{\zeta}_j(t): = \sum_{r=1}^{\infty} \gamma_{rj} e^{-i\omega_{rj} t} \hat{\mathcal{B}}_{rj}(0),
\end{align}
and $\hat{\tilde{\zeta}}_j(z)$ is Fourier-Laplace transform of $\hat{\zeta}_j(t)$.

The NESS is unique if $\det[\mathcal{M}(\omega)]=0$, for real $z=\omega$, has no solutions. It can be checked numerically that this is true for our set-up. When the NESS is unique, we can use Eq.\eqref{mat_eq} for real $z=\omega$ and neglect the terms proportional to $\hat{b}_j(0)$ and $\hat{c}_j(0)$. With real $z=\omega$, the Fourier-Laplace transform becomes a normal Fourier transform, and we can find NESS expectation values knowing the noise correlation functions in frequency space, which in turn, are governed by the initial state of the baths.
Below, we discuss how we use this to obtain our NESS results for both bosonic and fermionic systems consider the above setting.

\subsection{Bosonic setting: transmission in an input-output experiment}
\label{subsec: bosonic input output}

In the bosonic setting, the two additional leads can be treated as an input probe and an output probe. An input-output experiment then corresponds to the situation where a coherent drive, say at frequency $\omega_d$, is given to the input probe, and the intensity of the light at the output probe is measured. It can be modeled in our setting as follows. The initial state of the lead corresponding to the input probe is such that, modes corresponding to the specific frequency $\omega_d$ are in a coherent state, while all other modes are empty. All other baths, including the output probe are considered to be empty initially. The intensity of the output light is then given by the current into the output probe, which is proportional to the occupation number of the site with which the probe is attached.  This amounts to setting
\begin{align}
    & \langle\hat{\tilde{\zeta}}_l^\dagger(\omega)\hat{\tilde{\zeta}}_l(\omega') \rangle = (2\pi)^2 |\alpha|^2 \delta(\omega - \omega_d) \delta(\omega'- \omega_d),
\end{align}
where $|\alpha|$ is the strength of the coherent drive, and the index $l$ corresponds to the input probe, and similar noise correlation functions are for all other baths are zero. For left to right transmission, $\tau_{+}(\omega_d)$, the input probe is the left probe, so $l=1$, while for right to left transmission, $\tau_{-}(\omega_d)$, the input probe is the right probe, so, $l=N$.  
The left to right transmission is given by
\begin{align}
    \tau_{+}(\omega_d) :=\frac{\langle  \hat{c}_N^\dagger\hat{c}_N\rangle}{|\alpha|^2} = \Big | [\mathbf{G}(\omega_d)]_{N1} \Big|^2,~~{\rm with}~~l=1.
\end{align}
The right to left transmission is given by
\begin{align}
    \tau_{-}(\omega_d) :=\frac{\langle  \hat{c}_1^\dagger\hat{c}_1\rangle}{|\alpha|^2} = \Big | [\mathbf{G}(\omega_d)]_{1N} \Big|^2,~~{\rm with}~~l=N.
\end{align}

The above discussion can be generalized to probes attached at any two sites, which shows that, in general,  $\Big| [\mathbf{G}(\omega_d)]_{\ell j} \Big|^2$ is the transmission amplitude from site $j$ (input site) to site $\ell$ (output site).

\subsection{Fermionic setting: particle current}
\label{subsec: fermionic particle current}

In the fermionic setting, a chemical potential difference is given via the two additional leads to drive transport.

We consider the fermionic setting at inverse temperature $\beta$, and take the left lead to be at chemical potential $\mu_1$ and the right lead to be at chemical potential $\mu_N$. 
All local bulk baths are assumed to be empty (i.e., indicated by chemical potential $\to -\infty$) (see Fig.~\ref{fig. two_additional_leads}).
This amounts to using the noise correlation functions
\begin{align}
\label{fermionic_noise}
    \langle\hat{\tilde{\zeta}}_l^\dagger(\omega)\hat{\tilde{\zeta}}_l(\omega') \rangle = 2\pi \gamma n_l(\omega) \delta(\omega - \omega^\prime),~~l=1,N,
\end{align}
and similar noise correlation functions for all other baths are zero. Here $n_{l}(\omega)=[e^{\beta(\omega-\mu_l)}+1]^{-1}$ is the Fermi distribution function.

 In this set-up, we define the particle current into the right (left) lead $I_{S\to N}~(I_{S\to 1})$ as follows:
\begin{align}
    &I_{S\to N} = -i \left[ \hat{N}_{B_{N}}, \hat{\mathcal{H}}_{SB_N}\right] , \\
    &I_{S\to 1} = -i \left[ \hat{N}_{B_{1}}, \hat{\mathcal{H}}_{SB_1}\right],
\end{align}
where $\hat{N}_{B_{j}}=\sum_{r = 1}^{\infty} \hat{\mathcal{B}}_{rj}^{\dagger} \hat{\mathcal{B}}_{rj}$, $j=1,N$ is the total number operator for the additional leads. 
This gives,
\begin{align}
    &I_{S\to N}= i \langle\sum_{r}\gamma_{rN} \hat{c}_N^{\dagger} \hat{\mathcal{B}}_{rN}\rangle - i \langle\sum_{r}\gamma_{rN}^* \hat{\mathcal{B}}_{rN}^{\dagger}  \hat{c}_N\rangle ,\\
    & I_{S\to 1} = i \langle\sum_{r}\gamma_{r1} \hat{c}_1^{\dagger} \hat{\mathcal{B}}_{r1}\rangle - i \langle\sum_{r}\gamma_{r1}^* \hat{\mathcal{B}}_{r1}^{\dagger}  \hat{c}_1\rangle .
\end{align}
The currents $I_{\pm }(\mu_d)$, discussed in main text, are then given by
\begin{align}
    & I_{+}(\mu_d)=I_{S\to N},~~{\rm for}~~ \mu_1=\mu_d,~~\mu_N=-\infty,~~\beta\to \infty,\nonumber \\
    & I_{-}(\mu_d)=I_{S\to 1},~~{\rm for}~~ \mu_N=\mu_d,~~\mu_1=-\infty,~~\beta\to \infty.
\end{align}
The evolution of the bath operators, $\hat{B}_{rj}$, $j=1,N$, has the same form as Eq. \eqref{S10}. Utilizing this, performing Fourier transformation and using the noise correlation functions in Eq. \eqref{fermionic_noise}, we obtain the expressions for $I_{\pm }(\mu_d)$ as
\begin{align}
    I_{+}(\mu_d) =\gamma^2\int_{-\infty}^{\mu_d} \frac{d\omega}{2\pi} \Big | [\mathbf{G}(\omega)]_{N 1} \Big|^2 ,~~~I_{-}(\mu_d) =\gamma^2\int_{-\infty}^{\mu_d} \frac{d\omega}{2\pi} \Big | [\mathbf{G}(\omega)]_{1 N} \Big|^2 .
\end{align}

\section{Current scaling in Markovian fermionic case}
\label{Sec: Markovian fermionic current}

\begin{figure}
	\includegraphics[width=0.5\columnwidth]{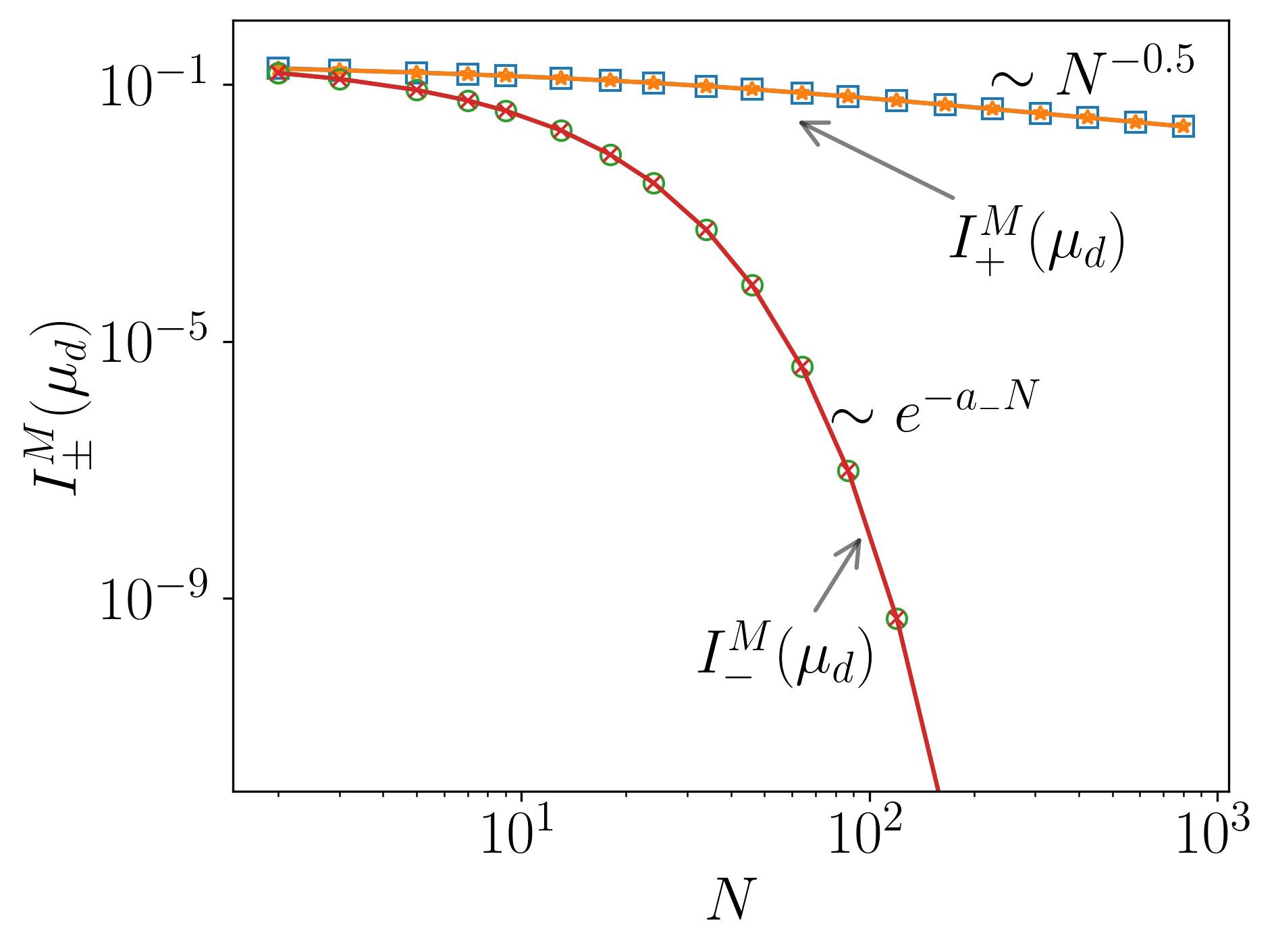}
	\caption{ \label{fig:Lyapunov_and_NEGF} The figure shows NESS current for Markovian case obtained from Lyapunov equation (open symbols) and from NEGF (filled symbols). Parameters: $\phi=2\pi/3, \omega_*=\Delta_c-g, \Gamma=(0.3)^2/g, \gamma=0.5g, \mu_d = \omega_*$ and inverse temperature set to $ \beta g =10$. As discussed, for all $\mu_d$, $I_+^M(\mu_d)$ decays as power-law with $N$, while $I_-^M(\mu_d)$ decays exponentially with $N$. There is no nonanalytic change with change in $\mu_d$. Hence no NDQPT is captured in the Markovian treatment.}
\end{figure}

The only way to drive transport via attaching two additional leads in the Markovian case, without changing the effective Hatano-Nelson model in the bulk, is via the so-called local GKSL approach, which gives the quantum master equation
\begin{align}
    \frac{\partial \hat{\rho}}{\partial t}= i[\hat{\rho},\hat{\mathcal{H}}_S]+ \sum_{j} \mathcal{D}^{L}_j[\hat{\rho}(t)] + \sum_{j=1,N}\mathcal{D}^{th}_j[\hat{\rho}(t)]
\end{align}
where 
\begin{align}
    &\mathcal{D}^{th}_j[\rho(t)] = \sum_{j=1,N} \Big(\gamma [1-n(\mu_j)]  ( \hat{c}_j \hat{\rho} \hat{c}_j^\dagger - \frac{1}{2}\{\hat{c}_j^\dagger \hat{c}_j, \hat{\rho} \})
    + \gamma n(\mu_j)( \hat{c}_j^\dagger \hat{\rho} \hat{c}_j - \frac{1}{2}\{\hat{c}_j \hat{c}_j^\dagger, \hat{\rho} \}) \Big),
\end{align}
and $\hat{\mathcal{H}}_S$ and $\mathcal{D}^{L}_j$ as defined in Eq.\eqref{H_NH_Lindblad}. The corresponding Lyapunov equation is
\begin{align}
\label{Lypunov_with_leads}
    &\frac{d\mathbf{C}}{dt} = i \mathbf{H}_{\rm eff}^\dagger
    \mathbf{C} - i\mathbf{C}\mathbf{H}_{\rm eff} + \mathbf{Q}, \nonumber \\
    &{\rm where}~~\mathbf{H}_{\rm eff}=\mathbf{H}_{\rm HN} +(\Delta_c - i\Gamma) \mathbb{I} - i \mathbf{Y},~~\mathbf{Y}_{11}=\mathbf{Y}_{NN}=\gamma,~~\mathbf{Q}_{11}=\gamma n(\mu_1),~~\mathbf{Q}_{NN}=\gamma n(\mu_N),   
\end{align}
all other elements of $\mathbf{Y}$ and $\mathbf{Q}$ are zero, and $\mathbf{H}_{\rm HN}$ is as given in Eq.\eqref{H_eff}. Under this equation,
\begin{align}
    I_{S\to N}= \gamma \langle \hat{c}_N^\dagger \hat{c}_N \rangle,~~I_{S\to 1}= \gamma \langle \hat{c}_1^\dagger \hat{c}_1 \rangle.
\end{align}
The steady state currents can be obtained by solving the Lyapunov equation for NESS, i.e, the algebraic Lyapunov equation
\begin{align}
    0= i \mathbf{H}_{\rm eff}^\dagger
    \mathbf{C} - i\mathbf{C}\mathbf{H}_{\rm eff} + \mathbf{Q}.
\end{align}
As in the non-Markovian case, the currents $I_{\pm }(\mu_d)$, discussed in main text, are then given by
\begin{align}
    & I_{+}(\mu_d)=I_{S\to N},~~{\rm for}~~ \mu_1=\mu_d,~~\mu_N=-\infty,~~\beta\to \infty,\nonumber \\
    & I_{-}(\mu_d)=I_{S\to 1},~~{\rm for}~~ \mu_N=\mu_d,~~\mu_1=-\infty,~~\beta\to \infty.
\end{align}
Using similar approach as in Appendix~\ref{subsec: HN Markovian NEGF}, it can be checked that following quantum Langevin equation leads to the same Lyapunov equation as Eq.\eqref{Lypunov_with_leads},
\begin{align}
\label{Markovian_QLE_with_leads}
    \frac{d\hat{c}(t)}{dt}=-i \mathbf{H}_{\rm eff}\hat{c}(t)-i \hat{\eta}(t)-i\hat{\zeta}(t),
\end{align}
where $\mathbf{H}_{\rm eff}$ is same as in Eq.\eqref{Lypunov_with_leads}, the elements of the column vector $\hat{\eta}(t)$ satisfy the same correlation functions as in Eq.\eqref{Markovian_QLE_noise}, and for the column vector $\hat{\zeta}(t)$, only first and $N$th elements $\hat{\zeta}_1(t)$ and $\hat{\zeta}_N(t)$ are non-zero, all other elements are zero. The operators $\hat{\zeta}_1(t)$ and $\hat{\zeta}_N(t)$ are noise operators due to the additional leads, and satisfy the correlation functions
\begin{align}
    & \langle \hat{\zeta}_1^\dagger(t) \hat{\zeta}_1(t^\prime) \rangle = \gamma n(\mu_1) \delta(t-t^\prime),~~
    \langle \hat{\zeta}_N^\dagger(t) \hat{\zeta}_N(t^\prime) \rangle = \gamma n(\mu_N)\delta(t-t^\prime),  \nonumber \\
    &\langle \hat{\zeta}_\ell(t)\rangle=0,~\langle \hat{\zeta}_1^\dagger(t) \hat{\zeta}_N(t^\prime) \rangle=0,~\langle \hat{\zeta}_j^\dagger(t) \hat{\eta}_\ell(t^\prime) \rangle =0,~~ \langle\hat{O}(0)\hat{\zeta}_\ell(t)\rangle=\langle\hat{\zeta}_\ell(t)\hat{O}(0)\rangle=0,
\end{align}
where $\hat{O}(0)$ is any system operator at initial time. The NESS currents can be obtained in terms of the retarded frequency-space NEGF by solving Eq.\eqref{Markovian_QLE_with_leads} via Fourier transform. This leads to
\begin{align}
    I_{+}^M(\mu_d) =\gamma^2 n(\mu_d)\int_{-\infty}^{\infty} \frac{d\omega}{2\pi} \Big | [\mathbf{G}^M(\omega)]_{N 1} \Big|^2 ,~~~I_{-}^M(\mu_d) =\gamma^2n(\mu_d)\int_{-\infty}^{\infty} \frac{d\omega}{2\pi} \Big | [\mathbf{G}^M(\omega)]_{1 N} \Big|^2 .
\end{align}
Due to the integration over the whole range of frequencies, the NDQPT on changing $\mu_d$ cannot be captured in the Markovian case. Instead, following steps akin to those in non-Markovian case, we obtain
\begin{align}
    I_{+}^M(\mu_d)\sim N^{-0.5}~~\forall~~\mu_d,~~~I_{-}^M(\mu_d)\sim e^{-a N}~~\forall~~\mu_d.
\end{align}
In Fig.~\ref{fig:Lyapunov_and_NEGF}, we show $I_{\pm}^M(\mu_d)$ obtained from both Lyapunov equation and from the NEGF for representative parameters to demonstrate that they perfectly match, and show the behavior mentioned above.

\twocolumngrid

\bibliography{ref}

@article{Gou_2020,
	title = {Tunable Nonreciprocal Quantum Transport through a Dissipative Aharonov-Bohm Ring in Ultracold Atoms},
	author = {Gou, Wei and Chen, Tao and Xie, Dizhou and Xiao, Teng and Deng, Tian-Shu and Gadway, Bryce and Yi, Wei and Yan, Bo},
	journal = {Phys. Rev. Lett.},
	volume = {124},
	issue = {7},
	pages = {070402},
	numpages = {6},
	year = {2020},
	month = {Feb},
	publisher = {American Physical Society},
	doi = {10.1103/PhysRevLett.124.070402},
	url = {https://link.aps.org/doi/10.1103/PhysRevLett.124.070402}
}

@article{Liang_2022,
	title = {Dynamic Signatures of Non-Hermitian Skin Effect and Topology in Ultracold Atoms},
	author = {Liang, Qian and Xie, Dizhou and Dong, Zhaoli and Li, Haowei and Li, Hang and Gadway, Bryce and Yi, Wei and Yan, Bo},
	journal = {Phys. Rev. Lett.},
	volume = {129},
	issue = {7},
	pages = {070401},
	numpages = {6},
	year = {2022},
	month = {Aug},
	publisher = {American Physical Society},
	doi = {10.1103/PhysRevLett.129.070401},
	url = {https://link.aps.org/doi/10.1103/PhysRevLett.129.070401}
}

@article{Ren_2023,
	author = {Chaojie Ren and Yuqing Li and Jizhou Wu and Hongxing Zhao and Yunfei Wang and Wenliang Liu and Peng Li and Yongming Fu and Liantuan Xiao and Jie Ma and Suotang Jia},
	journal = {Opt. Express},
	number = {21},
	pages = {34470--34476},
	publisher = {Optica Publishing Group},
	title = {Nonreciprocal dynamics of noninteracting ultracold atoms in a momentum lattice},
	volume = {31},
	month = {Oct},
	year = {2023},
	url = {https://opg.optica.org/oe/abstract.cfm?URI=oe-31-21-34470},
	doi = {10.1364/OE.500605}
}

@Article{Zhao_2025,
	author={Zhao, Entong
	and Wang, Zhiyuan
	and He, Chengdong
	and Poon, Ting Fung Jeffrey
	and Pak, Ka Kwan
	and Liu, Yu-Jun
	and Ren, Peng
	and Liu, Xiong-Jun
	and Jo, Gyu-Boong},
	title={Two-dimensional non-Hermitian skin effect in an ultracold Fermi gas},
	journal={Nature},
	year={2025},
	month={Jan},
	day={01},
	volume={637},
	number={8046},
	pages={565-573},
	issn={1476-4687},
	doi={10.1038/s41586-024-08347-3},
	url={https://doi.org/10.1038/s41586-024-08347-3}
}

@article{Wang_2024,
	author = {Ying-Ying Wang  and Yu-Xin Wang  and Sean van Geldern  and Thomas Connolly  and Aashish A. Clerk  and Chen Wang },
	title = {Dispersive nonreciprocity between a qubit and a cavity},
	journal = {Science Advances},
	volume = {10},
	number = {16},
	pages = {eadj8796},
	year = {2024},
	doi = {10.1126/sciadv.adj8796},
	URL = {https://www.science.org/doi/abs/10.1126/sciadv.adj8796}
	}

@article{Fedorov_2024,
	title = {Nonreciprocity and Circulation in a Passive Josephson-Junction Ring},
	author = {Fedorov, Arkady and Kumar, N. Pradeep and Le, Dat Thanh and Navarathna, Rohit and Pakkiam, Prasanna and Stace, Thomas M.},
	journal = {Phys. Rev. Lett.},
	volume = {132},
	issue = {9},
	pages = {097001},
	numpages = {6},
	year = {2024},
	month = {Feb},
	publisher = {American Physical Society},
	doi = {10.1103/PhysRevLett.132.097001},
	url = {https://link.aps.org/doi/10.1103/PhysRevLett.132.097001}
}

@Article{Ochkan_2024,
author={Ochkan, Kyrylo
and Chaturvedi, Raghav
and K{\"o}nye, Viktor
and Veyrat, Louis
and Giraud, Romain
and Mailly, Dominique
and Cavanna, Antonella
and Gennser, Ulf
and Hankiewicz, Ewelina M.
and B{\"u}chner, Bernd
and van den Brink, Jeroen
and Dufouleur, Joseph
and Fulga, Ion Cosma},
title={Non-Hermitian topology in a multi-terminal quantum Hall device},
journal={Nature Physics},
year={2024},
month={Mar},
day={01},
volume={20},
number={3},
pages={395-401},
issn={1745-2481},
doi={10.1038/s41567-023-02337-4},
url={https://doi.org/10.1038/s41567-023-02337-4}
}

@Article{Shen_2025,
	author={Shen, Ruizhe
	and Chen, Tianqi
	and Yang, Bo
	and Lee, Ching Hua},
	title={Observation of the non-Hermitian skin effect and Fermi skin on a digital quantum computer},
	journal={Nature Communications},
	year={2025},
	month={Feb},
	day={04},
	volume={16},
	number={1},
	pages={1340},
	issn={2041-1723},
	doi={10.1038/s41467-025-55953-4},
	url={https://doi.org/10.1038/s41467-025-55953-4}
}

@Article{Zhang_2025,
	author={Zhang, Yuxuan
	and Carrasquilla, Juan
	and Kim, Yong Baek},
	title={Observation of a non-Hermitian supersonic mode on a trapped-ion quantum computer},
	journal={Nature Communications},
	year={2025},
	month={Apr},
	day={06},
	volume={16},
	number={1},
	pages={3286},
	issn={2041-1723},
	doi={10.1038/s41467-025-57930-3},
	url={https://doi.org/10.1038/s41467-025-57930-3}
}

@article{Xiao_2021,
	title = {Observation of Non-Bloch Parity-Time Symmetry and Exceptional Points},
	author = {Xiao, Lei and Deng, Tianshu and Wang, Kunkun and Wang, Zhong and Yi, Wei and Xue, Peng},
	journal = {Phys. Rev. Lett.},
	volume = {126},
	issue = {23},
	pages = {230402},
	numpages = {6},
	year = {2021},
	month = {Jun},
	publisher = {American Physical Society},
	doi = {10.1103/PhysRevLett.126.230402},
	url = {https://link.aps.org/doi/10.1103/PhysRevLett.126.230402}
}

@Article{Xiao_2020,
	author={Xiao, Lei
	and Deng, Tianshu
	and Wang, Kunkun
	and Zhu, Gaoyan
	and Wang, Zhong
	and Yi, Wei
	and Xue, Peng},
	title={Non-Hermitian bulk--boundary correspondence in quantum dynamics},
	journal={Nature Physics},
	year={2020},
	month={Jul},
	day={01},
	volume={16},
	number={7},
	pages={761-766},
	issn={1745-2481},
	doi={10.1038/s41567-020-0836-6},
	url={https://doi.org/10.1038/s41567-020-0836-6}
}

@article{Metelmann_2015,
  title = {Nonreciprocal Photon Transmission and Amplification via Reservoir Engineering},
  author = {Metelmann, A. and Clerk, A. A.},
  journal = {Phys. Rev. X},
  volume = {5},
  issue = {2},
  pages = {021025},
  numpages = {16},
  year = {2015},
  month = {Jun},
  publisher = {American Physical Society},
  doi = {10.1103/PhysRevX.5.021025},
  url = {https://link.aps.org/doi/10.1103/PhysRevX.5.021025}
}

@article{McDonald_2022,
  title = {Nonequilibrium stationary states of quantum non-Hermitian lattice models},
  author = {McDonald, A. and Hanai, R. and Clerk, A. A.},
  journal = {Phys. Rev. B},
  volume = {105},
  issue = {6},
  pages = {064302},
  numpages = {19},
  year = {2022},
  month = {Feb},
  publisher = {American Physical Society},
  doi = {10.1103/PhysRevB.105.064302},
  url = {https://link.aps.org/doi/10.1103/PhysRevB.105.064302}
}

@Article{Clerk_2022,
	title={{Introduction to quantum non-reciprocal interactions: from non-Hermitian Hamiltonians to quantum master equations and quantum feedforward schemes}},
	author={Aashish A. Clerk},
	journal={SciPost Phys. Lect. Notes},
	pages={44},
	year={2022},
	publisher={SciPost},
	doi={10.21468/SciPostPhysLectNotes.44},
	url={https://scipost.org/10.21468/SciPostPhysLectNotes.44},
}

@article{Wang_2023,
  title = {Quantum Nonreciprocal Interactions via Dissipative Gauge Symmetry},
  author = {Wang, Yu-Xin and Wang, Chen and Clerk, Aashish A.},
  journal = {PRX Quantum},
  volume = {4},
  issue = {1},
  pages = {010306},
  numpages = {28},
  year = {2023},
  month = {Jan},
  publisher = {American Physical Society},
  doi = {10.1103/PRXQuantum.4.010306},
  url = {https://link.aps.org/doi/10.1103/PRXQuantum.4.010306}
}

@article{Begg_2024,
  title = {Quantum Criticality in Open Quantum Spin Chains with Nonreciprocity},
  author = {Begg, Samuel E. and Hanai, Ryo},
  journal = {Phys. Rev. Lett.},
  volume = {132},
  issue = {12},
  pages = {120401},
  numpages = {7},
  year = {2024},
  month = {Mar},
  publisher = {American Physical Society},
  doi = {10.1103/PhysRevLett.132.120401},
  url = {https://link.aps.org/doi/10.1103/PhysRevLett.132.120401}
}

@article{Belyansky_2025,
  title = {Phase Transitions in Nonreciprocal Driven-Dissipative Condensates},
  author = {Belyansky, Ron and Weis, Cheyne and Hanai, Ryo and Littlewood, Peter B. and Clerk, Aashish A.},
  journal = {Phys. Rev. Lett.},
  volume = {135},
  issue = {12},
  pages = {123401},
  numpages = {9},
  year = {2025},
  month = {Sep},
  publisher = {American Physical Society},
  doi = {10.1103/gphr-d1bc},
  url = {https://link.aps.org/doi/10.1103/gphr-d1bc}
}

@article{Soares_2025,
  title = {Dissipative Phase Transition of Interacting Nonreciprocal Fermions},
  author = {Soares, Rafael D. and Brunelli, Matteo and Schir\`o, Marco},
  journal = {Phys. Rev. Lett.},
  volume = {136},
  issue = {25},
  pages = {250403},
  numpages = {8},
  year = {2026},
  month = {Jun},
  publisher = {American Physical Society},
  doi = {10.1103/n8xv-kygx},
  url = {https://link.aps.org/doi/10.1103/n8xv-kygx}
}

@article{Brighi_2025,
      title={Pairing-induced phase transition in the non-reciprocal Kitaev chain}, 
      author={Pietro Brighi and Andreas Nunnenkamp},
      year={2025},
      eprint={2510.24851},
      archivePrefix={arXiv},
      primaryClass={quant-ph},
      url={https://arxiv.org/abs/2510.24851}, 
}

@article{Yi_2025,
      title={Nonreciprocity and unidirectional invisibility in three optical modes with non-Markovian effects}, 
      author={H. Yi and T. Z. Luan and W. Y. Hu and Cheng Shang and Yan-Hui Zhou and Zhi-Cheng Shi and H. Z. Shen},
      year={2025},
      eprint={2503.23169},
      archivePrefix={arXiv},
      primaryClass={physics.optics},
      url={https://arxiv.org/abs/2503.23169}, 
}

@article{luan_2025,
      title={Nonreciprocal quantum router with non-Markovian environments}, 
      author={T. Z. Luan and Cheng Shang and H. Yi and J. L. Li and Yan-Hui Zhou and Shuang Xu and H. Z. Shen},
      year={2025},
      eprint={2503.18647},
      archivePrefix={arXiv},
      primaryClass={physics.optics},
      url={https://arxiv.org/abs/2503.18647}, 
}

@article{Kuo_2025,
  title = {Non-Markovian skin effect},
  author = {Kuo, Po-Chen and Yang, Shen-Liang and Lambert, Neill and Lin, Jhen-Dong and Huang, Yi-Te and Nori, Franco and Chen, Yueh-Nan},
  journal = {Phys. Rev. Res.},
  volume = {7},
  issue = {1},
  pages = {L012068},
  numpages = {7},
  year = {2025},
  month = {Mar},
  publisher = {American Physical Society},
  doi = {10.1103/PhysRevResearch.7.L012068},
  url = {https://link.aps.org/doi/10.1103/PhysRevResearch.7.L012068}
}

@Article{Wilkey_2023_1,
author={Wilkey, Andrew
and Suelzer, Joseph
and Joglekar, Yogesh N.
and Vemuri, Gautam},
title={Theoretical and experimental characterization of non-Markovian anti-parity-time systems},
journal={Communications Physics},
year={2023},
month={Oct},
day={20},
volume={6},
number={1},
pages={308},
issn={2399-3650},
doi={10.1038/s42005-023-01426-3},
url={https://doi.org/10.1038/s42005-023-01426-3}
}

@Article{Wilkey_2023_2,
AUTHOR = {Wilkey, Andrew and Joglekar, Yogesh N. and Vemuri, Gautam},
TITLE = {Exceptional Points in a Non-Markovian Anti-Parity-Time Symmetric System},
JOURNAL = {Photonics},
VOLUME = {10},
YEAR = {2023},
NUMBER = {12},
ARTICLE-NUMBER = {1299},
URL = {https://www.mdpi.com/2304-6732/10/12/1299},
ISSN = {2304-6732},
DOI = {10.3390/photonics10121299}
}

@article{Wu_2026,
  title = {Quantum Mpemba Effect Induced by Non-Markovian Exceptional Points},
  author = {Zhang, Ze-Zhou and Luo, Hong-Gang and Wu, Wei},
  journal = {Phys. Rev. Lett.},
  volume = {136},
  issue = {21},
  pages = {210402},
  numpages = {8},
  year = {2026},
  month = {May},
  publisher = {American Physical Society},
  doi = {10.1103/zjdz-rqqd},
  url = {https://link.aps.org/doi/10.1103/zjdz-rqqd}
}

@article{Molina_2026,
      title={Bath memory as a precision resource in quantum transport}, 
      author={José Molina and Sheikh Parvez Mandal and Mahasweta Pandit and Javier Prior},
      year={2026},
      eprint={2606.17026},
      archivePrefix={arXiv},
      primaryClass={quant-ph},
      url={https://arxiv.org/abs/2606.17026}, 
}

@Article{Huang_2026,
author={Huang, Juntao
and Hu, Jiangping
and Yang, Zhesen},
title={Complex frequency detection in a subsystem},
journal={Communications Physics},
year={2026},
month={Feb},
day={03},
volume={9},
number={1},
pages={84},
issn={2399-3650},
doi={10.1038/s42005-026-02524-8},
url={https://doi.org/10.1038/s42005-026-02524-8}
}

@article{Dordevic_2026,
  title = {Non-Hermitian Crystalline Braid Topology from Hermitian Projection: A Zero-Mode Resonance Mechanism},
  author = {Stefan Djordjevi\'{c} and Vladimir Juri\v{c}i\'{c}},
  year = {2026},
  eprint = {2606.06626},
  archivePrefix = {arXiv},
  primaryClass = {cond-mat.mes-hall},
}

@article{Ashida_2020,
  title={Non-hermitian physics},
  author={Ashida, Yuto and Gong, Zongping and Ueda, Masahito},
  journal={Advances in Physics},
  volume={69},
  number={3},
  pages={249--435},
  year={2020},
  publisher={Taylor \& Francis},
  doi={10.1080/00018732.2021.1876991},
  url={https://doi.org/10.1080/00018732.2021.1876991}
}

@article{Yang_2023,
author = {Changqing Wang and Zhoutian Fu and Wenbo Mao and Jinran Qie and A. Douglas Stone and Lan Yang},
journal = {Adv. Opt. Photon.},
number = {2},
pages = {442--523},
publisher = {Optica Publishing Group},
title = {Non-Hermitian optics and photonics: from classical to quantum},
volume = {15},
month = {Jun},
year = {2023},
url = {https://doi.org/10.1364/AOP.475477},
doi = {10.1364/AOP.475477}
}

@article{Zhang_2018,
doi = {10.1088/1361-6455/aaaf9f},
url = {https://doi.org/10.1088/1361-6455/aaaf9f},
year = {2018},
month = {mar},
publisher = {IOP Publishing},
volume = {51},
number = {7},
pages = {072001},
author = {Zhang, Zhaoyang and Ma, Danmeng and Sheng, Jiteng and Zhang, Yiqi and Zhang, Yanpeng and Xiao, Min},
title = {Non-Hermitian optics in atomic systems},
journal = {Journal of Physics B: Atomic, Molecular and Optical Physics}
}

@Article{Lin_2025,
author={Lin, Jhen-Dong
and Kuo, Po-Chen
and Lambert, Neill
and Miranowicz, Adam
and Nori, Franco
and Chen, Yueh-Nan},
title={Non-Markovian quantum exceptional points},
journal={Nature Communications},
year={2025},
month={Feb},
day={03},
volume={16},
number={1},
pages={1289},
issn={2041-1723},
doi={10.1038/s41467-025-56242-w},
url={https://doi.org/10.1038/s41467-025-56242-w}
}

@article{Longhi_2016,
  title = {Non-Hermitian tight-binding network engineering},
  author = {Longhi, Stefano},
  journal = {Phys. Rev. A},
  volume = {93},
  issue = {2},
  pages = {022102},
  numpages = {9},
  year = {2016},
  month = {Feb},
  publisher = {American Physical Society},
  doi = {10.1103/PhysRevA.93.022102},
  url = {https://link.aps.org/doi/10.1103/PhysRevA.93.022102}
}

@article{Rotter_2009,
doi = {10.1088/1751-8113/42/15/153001},
url = {https://doi.org/10.1088/1751-8113/42/15/153001},
year = {2009},
month = {mar},
publisher = {},
volume = {42},
number = {15},
pages = {153001},
author = {Rotter, Ingrid},
title = {A non-Hermitian Hamilton operator and the physics of open quantum systems},
journal = {Journal of Physics A: Mathematical and Theoretical},
}

@article{Ordonez_2014,
    author = {Hatano, Naomichi and Ordonez, Gonzalo},
    title = {Time-reversal symmetric resolution of unity without background integrals in open quantum systems},
    journal = {Journal of Mathematical Physics},
    volume = {55},
    number = {12},
    pages = {122106},
    year = {2014},
    month = {12},
    issn = {0022-2488},
    doi = {10.1063/1.4904200},
    url = {https://doi.org/10.1063/1.4904200},
}

@article{Ordonez_2017,
doi = {10.1088/1751-8121/aa85ae},
url = {https://doi.org/10.1088/1751-8121/aa85ae},
year = {2017},
month = {sep},
publisher = {IOP Publishing},
volume = {50},
number = {40},
pages = {405304},
author = {Ordonez, Gonzalo and Hatano, Naomichi},
title = {The arrow of time in open quantum systems and dynamical breaking of the resonance–anti-resonance symmetry},
journal = {Journal of Physics A: Mathematical and Theoretical},

}

@article{Zhang_2_2022,
author = {Xiujuan Zhang and Tian Zhang and Ming-Hui Lu and Yan-Feng Chen},
title = {A review on non-Hermitian skin effect},
journal = {Advances in Physics: X},
volume = {7},
number = {1},
pages = {2109431},
year = {2022},
publisher = {Taylor \& Francis},
doi = {10.1080/23746149.2022.2109431},
URL = {https://doi.org/10.1080/23746149.2022.2109431},
}

@article{Nagaosa_2024,
   author = "Nagaosa, Naoto and Yanase, Youichi",
   title = "Nonreciprocal Transport and Optical Phenomena in Quantum Materials", 
   journal= "Annual Review of Condensed Matter Physics",
   year = "2024",
   volume = "15",
   number = "Volume 15, 2024",
   pages = "63-83",
   doi = "https://doi.org/10.1146/annurev-conmatphys-032822-033734",
   url = "https://www.annualreviews.org/content/journals/10.1146/annurev-conmatphys-032822-033734",
   publisher = "Annual Reviews",
   issn = "1947-5462",
   type = "Journal Article"
  }

@Article{Ding_2022,
author={Ding, Kun
and Fang, Chen
and Ma, Guancong},
title={Non-Hermitian topology and exceptional-point geometries},
journal={Nature Reviews Physics},
year={2022},
month={Dec},
day={01},
volume={4},
number={12},
pages={745-760},
issn={2522-5820},
doi={10.1038/s42254-022-00516-5},
url={https://doi.org/10.1038/s42254-022-00516-5}
}

@article{Wang_2_2023,
author = {Aoxi Wang  and Zhiqiang Meng  and Chang Qing Chen },
title = {Non-Hermitian topology in static mechanical metamaterials},
journal = {Science Advances},
volume = {9},
number = {27},
pages = {eadf7299},
year = {2023},
doi = {10.1126/sciadv.adf7299},
URL = {https://www.science.org/doi/abs/10.1126/sciadv.adf7299},
}

@article{Zhu_2024,
doi = {10.1088/1361-648X/ad3593},
url = {https://doi.org/10.1088/1361-648X/ad3593},
year = {2024},
month = {mar},
publisher = {IOP Publishing},
volume = {36},
number = {25},
pages = {253003},
author = {Zhu, Weiwei and Li, Linhu},
title = {A brief review of hybrid skin-topological effect},
journal = {Journal of Physics: Condensed Matter}
}

@article{Fruchart_2021,
  title={Non-reciprocal phase transitions},
  author={Fruchart, Michel and Hanai, Ryo and Littlewood, Peter B and Vitelli, Vincenzo},
  journal={Nature},
  volume={592},
  number={7854},
  pages={363--369},
  year={2021},
  publisher={Nature Publishing Group UK London},
  doi={10.1038/s41586-021-03375-9},
  url={https://www.nature.com/articles/s41586-021-03375-9}
}

@article{Lau_2018,
  title={Fundamental limits and non-reciprocal approaches in non-Hermitian quantum sensing},
  author={Lau, Hoi-Kwan and Clerk, Aashish A},
  journal={Nature communications},
  volume={9},
  number={1},
  pages={4320},
  year={2018},
  publisher={Nature Publishing Group UK London},
  doi={10.1038/s41467-018-06477-7},
  url={https://www.nature.com/articles/s41467-018-06477-7}
}

@article{Bao_2021,
  title = {Fundamental limits for reciprocal and nonreciprocal non-Hermitian quantum sensing},
  author = {Bao, Liying and Qi, Bo and Dong, Daoyi and Nori, Franco},
  journal = {Phys. Rev. A},
  volume = {103},
  issue = {4},
  pages = {042418},
  numpages = {13},
  year = {2021},
  month = {Apr},
  publisher = {American Physical Society},
  doi = {10.1103/PhysRevA.103.042418},
  url = {https://link.aps.org/doi/10.1103/PhysRevA.103.042418}
}

@article{Xie_2024,
  title = {Quantum sensing with nonreciprocal couplings},
  author = {Xie, Dong and Xu, Chunling},
  journal = {Phys. Rev. Appl.},
  volume = {22},
  issue = {6},
  pages = {064072},
  numpages = {12},
  year = {2024},
  month = {Dec},
  publisher = {American Physical Society},
  doi = {10.1103/PhysRevApplied.22.064072},
  url = {https://link.aps.org/doi/10.1103/PhysRevApplied.22.064072}
}

@article{Brighi_2024,
  title = {Nonreciprocal dynamics and the non-Hermitian skin effect of repulsively bound pairs},
  author = {Brighi, Pietro and Nunnenkamp, Andreas},
  journal = {Phys. Rev. A},
  volume = {110},
  issue = {2},
  pages = {L020201},
  numpages = {6},
  year = {2024},
  month = {Aug},
  publisher = {American Physical Society},
  doi = {10.1103/PhysRevA.110.L020201},
  url = {https://link.aps.org/doi/10.1103/PhysRevA.110.L020201}
}

@Article{McDonald_2020,
author={McDonald, Alexander
and Clerk, Aashish A.},
title={Exponentially-enhanced quantum sensing with non-Hermitian lattice dynamics},
journal={Nature Communications},
year={2020},
month={Oct},
day={23},
volume={11},
number={1},
pages={5382},
doi={10.1038/s41467-020-19090-4},
url={https://doi.org/10.1038/s41467-020-19090-4}
}

@article{GKSL_1976,
    author = {Gorini, Vittorio and Kossakowski, Andrzej and Sudarshan, E. C. G.},
    title = {Completely positive dynamical semigroups of N‐level systems},
    journal = {Journal of Mathematical Physics},
    volume = {17},
    number = {5},
    pages = {821--825},
    year = {1976},
    month = {05},
    issn = {0022-2488},
    doi = {10.1063/1.522979},
    url = {https://doi.org/10.1063/1.522979}
}

@article{Lindblad1976,
author={Lindblad, G\"{o}ran},
title = {On the generators of quantum dynamical semigroups},
journal = {Communications in Mathematical Physics},
    volume = {48},
    number = {2},
    pages = {119--130},
    year = {1976},
    month = {06},
     doi = {https://doi.org/10.1007/BF01608499},
    url = {https://doi.org/10.1007/BF01608499}
}

@article{Breuer_2016,
  title = {Colloquium: Non-Markovian dynamics in open quantum systems},
  author = {Breuer, Heinz-Peter and Laine, Elsi-Mari and Piilo, Jyrki and Vacchini, Bassano},
  journal = {Rev. Mod. Phys.},
  volume = {88},
  issue = {2},
  pages = {021002},
  numpages = {24},
  year = {2016},
  month = {Apr},
  publisher = {American Physical Society},
  doi = {10.1103/RevModPhys.88.021002},
  url = {https://link.aps.org/doi/10.1103/RevModPhys.88.021002}
}

@article{Vega_2017,
  title = {Dynamics of non-Markovian open quantum systems},
  author = {de Vega, In\'es and Alonso, Daniel},
  journal = {Rev. Mod. Phys.},
  volume = {89},
  issue = {1},
  pages = {015001},
  numpages = {58},
  year = {2017},
  month = {Jan},
  publisher = {American Physical Society},
  doi = {10.1103/RevModPhys.89.015001},
  url = {https://link.aps.org/doi/10.1103/RevModPhys.89.015001}
}

@article{Cialdi_2019,
  title = {Experimental investigation of the effect of classical noise on quantum non-Markovian dynamics},
  author = {Cialdi, Simone and Benedetti, Claudia and Tamascelli, Dario and Olivares, Stefano and Paris, Matteo G. A. and Vacchini, Bassano},
  journal = {Phys. Rev. A},
  volume = {100},
  issue = {5},
  pages = {052104},
  numpages = {7},
  year = {2019},
  month = {Nov},
  publisher = {American Physical Society},
  doi = {10.1103/PhysRevA.100.052104},
  url = {https://link.aps.org/doi/10.1103/PhysRevA.100.052104}
}

@article{Xu_2010,
  title = {Experimental investigation of the non-Markovian dynamics of classical and quantum correlations},
  author = {Xu, Jin-Shi and Li, Chuan-Feng and Zhang, Cheng-Jie and Xu, Xiao-Ye and Zhang, Yong-Sheng and Guo, Guang-Can},
  journal = {Phys. Rev. A},
  volume = {82},
  issue = {4},
  pages = {042328},
  numpages = {7},
  year = {2010},
  month = {Oct},
  publisher = {American Physical Society},
  doi = {10.1103/PhysRevA.82.042328},
  url = {https://link.aps.org/doi/10.1103/PhysRevA.82.042328}
}

@Article{Liu_2011,
author={Liu, Bi-Heng
and Li, Li
and Huang, Yun-Feng
and Li, Chuan-Feng
and Guo, Guang-Can
and Laine, Elsi-Mari
and Breuer, Heinz-Peter
and Piilo, Jyrki},
title={Experimental control of the transition from Markovian to non-Markovian dynamics of open quantum systems},
journal={Nature Physics},
year={2011},
month={Dec},
day={01},
volume={7},
number={12},
pages={931-934},
issn={1745-2481},
doi={10.1038/nphys2085},
url={https://doi.org/10.1038/nphys2085}
}

@Article{Bernardes_2015,
author={Bernardes, Nadja K.
and Cuevas, Alvaro
and Orieux, Adeline
and Monken, C. H.
and Mataloni, Paolo
and Sciarrino, Fabio
and Santos, Marcelo F.},
title={Experimental observation of weak non-Markovianity},
journal={Scientific Reports},
year={2015},
month={Dec},
day={02},
volume={5},
number={1},
pages={17520},
issn={2045-2322},
doi={10.1038/srep17520},
url={https://doi.org/10.1038/srep17520}
}

@article{Ramos_2016,
  title = {Non-Markovian dynamics in chiral quantum networks with spins and photons},
  author = {Ramos, Tom\'as and Vermersch, Beno\^{\i}t and Hauke, Philipp and Pichler, Hannes and Zoller, Peter},
  journal = {Phys. Rev. A},
  volume = {93},
  issue = {6},
  pages = {062104},
  numpages = {23},
  year = {2016},
  month = {Jun},
  publisher = {American Physical Society},
  doi = {10.1103/PhysRevA.93.062104},
  url = {https://link.aps.org/doi/10.1103/PhysRevA.93.062104}
}

@Article{White_2020,
author={White, G. A. L.
and Hill, C. D.
and Pollock, F. A.
and Hollenberg, L. C. L.
and Modi, K.},
title={Demonstration of non-Markovian process characterisation and control on a quantum processor},
journal={Nature Communications},
year={2020},
month={Dec},
day={09},
volume={11},
number={1},
pages={6301},
issn={2041-1723},
doi={10.1038/s41467-020-20113-3},
url={https://doi.org/10.1038/s41467-020-20113-3}
}

@article{Uriri_2020,
  title = {Experimental investigation of Markovian and non-Markovian channel addition},
  author = {Uriri, S. A. and Wudarski, F. and Sinayskiy, I. and Petruccione, F. and Tame, M. S.},
  journal = {Phys. Rev. A},
  volume = {101},
  issue = {5},
  pages = {052107},
  numpages = {10},
  year = {2020},
  month = {May},
  publisher = {American Physical Society},
  doi = {10.1103/PhysRevA.101.052107},
  url = {https://link.aps.org/doi/10.1103/PhysRevA.101.052107}
}

@article{Guo_2021,
  title = {Experimental Demonstration of Instrument-Specific Quantum Memory Effects and Non-Markovian Process Recovery for Common-Cause Processes},
  author = {Guo, Yu and Taranto, Philip and Liu, Bi-Heng and Hu, Xiao-Min and Huang, Yun-Feng and Li, Chuan-Feng and Guo, Guang-Can},
  journal = {Phys. Rev. Lett.},
  volume = {126},
  issue = {23},
  pages = {230401},
  numpages = {6},
  year = {2021},
  month = {Jun},
  publisher = {American Physical Society},
  doi = {10.1103/PhysRevLett.126.230401},
  url = {https://link.aps.org/doi/10.1103/PhysRevLett.126.230401}
}

@article{Goswami_2021,
  title = {Experimental characterization of a non-Markovian quantum process},
  author = {Goswami, K. and Giarmatzi, C. and Monterola, C. and Shrapnel, S. and Romero, J. and Costa, F.},
  journal = {Phys. Rev. A},
  volume = {104},
  issue = {2},
  pages = {022432},
  numpages = {7},
  year = {2021},
  month = {Aug},
  publisher = {American Physical Society},
  doi = {10.1103/PhysRevA.104.022432},
  url = {https://link.aps.org/doi/10.1103/PhysRevA.104.022432}
}

@article{Gaikwad_2024,
  title = {Entanglement Assisted Probe of the Non-Markovian to Markovian Transition in Open Quantum System Dynamics},
  author = {Gaikwad, Chandrashekhar and Kowsari, Daria and Brame, Carson and Song, Xingrui and Zhang, Haimeng and Esposito, Martina and Ranadive, Arpit and Cappelli, Giulio and Roch, Nicolas and Levenson-Falk, Eli M. and Murch, Kater W.},
  journal = {Phys. Rev. Lett.},
  volume = {132},
  issue = {20},
  pages = {200401},
  numpages = {7},
  year = {2024},
  month = {May},
  publisher = {American Physical Society},
  doi = {10.1103/PhysRevLett.132.200401},
  url = {https://link.aps.org/doi/10.1103/PhysRevLett.132.200401}
}

@article{Agarwal_2024,
doi = {10.1088/2058-9565/ad3d7e},
url = {https://doi.org/10.1088/2058-9565/ad3d7e},
year = {2024},
month = {apr},
publisher = {IOP Publishing},
volume = {9},
number = {3},
pages = {035017},
author = {Agarwal, Abhishek and Lindoy, Lachlan P and Lall, Deep and Jamet, François and Rungger, Ivan},
title = {Modelling non-Markovian noise in driven superconducting qubits},
journal = {Quantum Science and Technology}
}

@Article{Odeh_2025,
author={Odeh, Mutasem
and Godeneli, Kadircan
and Li, Eric
and Tangirala, Rohin
and Zhou, Haoxin
and Zhang, Xueyue
and Zhang, Zi-Huai
and Sipahigil, Alp},
title={Non-Markovian dynamics of a superconducting qubit in a phononic bandgap},
journal={Nature Physics},
year={2025},
month={Mar},
day={01},
volume={21},
number={3},
pages={406-411},
issn={1745-2481},
doi={10.1038/s41567-024-02740-5},
url={https://doi.org/10.1038/s41567-024-02740-5}
}

@Article{Fang_2017,
author={Fang, Kejie
and Luo, Jie
and Metelmann, Anja
and Matheny, Matthew H.
and Marquardt, Florian
and Clerk, Aashish A.
and Painter, Oskar},
title={Generalized non-reciprocity in an optomechanical circuit via synthetic magnetism and reservoir engineering},
journal={Nature Physics},
year={2017},
month={May},
day={01},
volume={13},
number={5},
pages={465-471},
issn={1745-2481},
doi={10.1038/nphys4009},
url={https://doi.org/10.1038/nphys4009}
}

@article{Hatano_1997,
  title = {Vortex pinning and non-Hermitian quantum mechanics},
  author = {Hatano, Naomichi and Nelson, David R.},
  journal = {Phys. Rev. B},
  volume = {56},
  issue = {14},
  pages = {8651--8673},
  numpages = {0},
  year = {1997},
  month = {Oct},
  publisher = {American Physical Society},
  doi = {10.1103/PhysRevB.56.8651},
  url = {https://link.aps.org/doi/10.1103/PhysRevB.56.8651}
}

@article{Hatano_1996,
  title = {Localization Transitions in Non-Hermitian Quantum Mechanics},
  author = {Hatano, Naomichi and Nelson, David R.},
  journal = {Phys. Rev. Lett.},
  volume = {77},
  issue = {3},
  pages = {570--573},
  numpages = {0},
  year = {1996},
  month = {Jul},
  publisher = {American Physical Society},
  doi = {10.1103/PhysRevLett.77.570},
  url = {https://link.aps.org/doi/10.1103/PhysRevLett.77.570}
}

@article{Liu_2021,
  title={Non-Hermitian skin effect in a non-Hermitian electrical circuit},
  author={Liu, Shuo and Shao, Ruiwen and Ma, Shaojie and Zhang, Lei and You, Oubo and Wu, Haotian and Xiang, Yuan Jiang and Cui, Tie Jun and Zhang, Shuang},
  journal={Research},
  year={2021},
  publisher={AAAS},
  doi={10.34133/2021/5608038},
  url={https://doi.org/10.34133/2021/560803}
}

@Article{Helbig_2020,
author={Helbig, T.
and Hofmann, T.
and Imhof, S.
and Abdelghany, M.
and Kiessling, T.
and Molenkamp, L. W.
and Lee, C. H.
and Szameit, A.
and Greiter, M.
and Thomale, R.},
title={Generalized bulk--boundary correspondence in non-Hermitian topolectrical circuits},
journal={Nature Physics},
year={2020},
month={Jul},
day={01},
volume={16},
number={7},
pages={747-750},
issn={1745-2481},
doi={10.1038/s41567-020-0922-9},
url={https://doi.org/10.1038/s41567-020-0922-9}
}

@Article{Brandenbourger_2019,
author={Brandenbourger, Martin
and Locsin, Xander
and Lerner, Edan
and Coulais, Corentin},
title={Non-reciprocal robotic metamaterials},
journal={Nature Communications},
year={2019},
month={Oct},
day={10},
volume={10},
number={1},
pages={4608},
issn={2041-1723},
doi={10.1038/s41467-019-12599-3},
url={https://doi.org/10.1038/s41467-019-12599-3}
}

@Article{Lianchao_2023,
author={Wang, Lianchao
and Mart{\'i}nez, Julio A. Iglesias
and Ulliac, Gwenn
and Wang, Bing
and Laude, Vincent
and Kadic, Muamer},
title={Non-reciprocal and non-Newtonian mechanical metamaterials},
journal={Nature Communications},
year={2023},
month={Aug},
day={08},
volume={14},
number={1},
pages={4778},
issn={2041-1723},
doi={10.1038/s41467-023-40493-6},
url={https://doi.org/10.1038/s41467-023-40493-6}
}

@Article{Veenstra_2024,
author={Veenstra, Jonas
and Gamayun, Oleksandr
and Guo, Xiaofei
and Sarvi, Anahita
and Meinersen, Chris Ventura
and Coulais, Corentin},
title={Non-reciprocal topological solitons in active metamaterials},
journal={Nature},
year={2024},
month={Mar},
day={01},
volume={627},
number={8004},
pages={528-533},
issn={1476-4687},
doi={10.1038/s41586-024-07097-6},
url={https://doi.org/10.1038/s41586-024-07097-6}
}

@Article{Yang_2024,
	author={Yang, Yang
	and Chapman, Robert J.
	and Haylock, Ben
	and Lenzini, Francesco
	and Joglekar, Yogesh N.
	and Lobino, Mirko
	and Peruzzo, Alberto},
	title={Programmable high-dimensional Hamiltonian in a photonic waveguide array},
	journal={Nature Communications},
	year={2024},
	month={Jan},
	day={02},
	volume={15},
	number={1},
	pages={50},
	issn={2041-1723},
	doi={10.1038/s41467-023-44185-z},
	url={https://doi.org/10.1038/s41467-023-44185-z}
}

@article{Wang_24,
author = {Xilan Wang and Ran Hao and Pengtao Fan and Luoshu Hu and Bilin Ye and Yonggang Zou and Shangzhong Jin},
journal = {Opt. Lett.},
number = {3},
pages = {554--557},
publisher = {Optica Publishing Group},
title = {Effective enhancement of the non-Hermitian corner skin effect in reciprocal photonic crystals},
volume = {49},
month = {Feb},
year = {2024},
url = {https://doi.org/10.1364/OL.513800},
doi = {10.1364/OL.513800},
}

@Article{Popa_2014,
author={Popa, Bogdan-Ioan
and Cummer, Steven A.},
title={Non-reciprocal and highly nonlinear active acoustic metamaterials},
journal={Nature Communications},
year={2014},
month={Feb},
day={27},
volume={5},
number={1},
pages={3398},
issn={2041-1723},
doi={10.1038/ncomms4398},
url={https://doi.org/10.1038/ncomms4398}
}

@Article{Shao_2020,
author={Shao, Linbo
and Mao, Wenbo
and Maity, Smarak
and Sinclair, Neil
and Hu, Yaowen
and Yang, Lan
and Lon{\v{c}}ar, Marko},
title={Non-reciprocal transmission of microwave acoustic waves in nonlinear parity--time symmetric resonators},
journal={Nature Electronics},
year={2020},
month={May},
day={01},
volume={3},
number={5},
pages={267-272},
issn={2520-1131},
doi={10.1038/s41928-020-0414-z},
url={https://doi.org/10.1038/s41928-020-0414-z}
}

@article{rasmussen_2021,
  title={Acoustic nonreciprocity},
  author={Rasmussen, Curtis and Quan, Li and Al{\`u}, Andrea},
  journal={Journal of Applied Physics},
  volume={129},
  number={21},
  year={2021},
  publisher={AIP Publishing},
  doi={10.1063/5.0050775},
  url={https://doi.org/10.1063/5.0050775}
}

@article{Molinari_1998,
doi = {10.1088/0305-4470/31/42/014},
url = {https://doi.org/10.1088/0305-4470/31/42/014},
year = {1998},
month = {oct},
publisher = {},
volume = {31},
number = {42},
pages = {8553},
author = {Luca Molinari},
title = {Transfer matrices, non-Hermitian Hamiltonians and resolvents: some spectral identities}
}

@article{Molinari_1997,
doi = {10.1088/0305-4470/30/3/021},
url = {https://doi.org/10.1088/0305-4470/30/3/021},
year = {1997},
month = {feb},
publisher = {},
volume = {30},
number = {3},
pages = {983},
author = {Luca Molinari},
title = {Transfer matrices and tridiagonal-block Hamiltonians with periodic and scattering boundary conditions},
journal = {Journal of Physics A: Mathematical and General}
}

@article{Lavis_1997,
title = {The inverse of a symmetric banded toeplitz matrix},
journal = {Reports on Mathematical Physics},
volume = {39},
number = {1},
pages = {137-146},
year = {1997},
issn = {0034-4877},
doi = {https://doi.org/10.1016/S0034-4877(97)81478-4},
url = {https://www.sciencedirect.com/science/article/pii/S0034487797814784},
author = {D.A. Lavis and B.W. Southern}
}

@article{Dwivedi_2016,
  title = {Of bulk and boundaries: Generalized transfer matrices for tight-binding models},
  author = {Dwivedi, Vatsal and Chua, Victor},
  journal = {Phys. Rev. B},
  volume = {93},
  issue = {13},
  pages = {134304},
  numpages = {26},
  year = {2016},
  month = {Apr},
  publisher = {American Physical Society},
  doi = {10.1103/PhysRevB.93.134304},
  url = {https://link.aps.org/doi/10.1103/PhysRevB.93.134304}
}

@article{Kunst_2019,
  title = {Non-Hermitian systems and topology: A transfer-matrix perspective},
  author = {Kunst, Flore K. and Dwivedi, Vatsal},
  journal = {Phys. Rev. B},
  volume = {99},
  issue = {24},
  pages = {245116},
  numpages = {21},
  year = {2019},
  month = {Jun},
  publisher = {American Physical Society},
  doi = {10.1103/PhysRevB.99.245116},
  url = {https://link.aps.org/doi/10.1103/PhysRevB.99.245116}
}

@Article{Mills_2019,
author={Mills, A. R.
and Zajac, D. M.
and Gullans, M. J.
and Schupp, F. J.
and Hazard, T. M.
and Petta, J. R.},
title={Shuttling a single charge across a one-dimensional array of silicon quantum dots},
journal={Nature Communications},
year={2019},
month={Mar},
day={05},
volume={10},
number={1},
pages={1063},
issn={2041-1723},
doi={10.1038/s41467-019-08970-z},
url={https://doi.org/10.1038/s41467-019-08970-z}
}

@Article{Borsoi_2024,
author={Borsoi, Francesco
and Hendrickx, Nico W.
and John, Valentin
and Meyer, Marcel
and Motz, Sayr
and van Riggelen, Floor
and Sammak, Amir
and de Snoo, Sander L.
and Scappucci, Giordano
and Veldhorst, Menno},
title={Shared control of a 16{\thinspace}semiconductor quantum dot crossbar array},
journal={Nature Nanotechnology},
year={2024},
month={Jan},
day={01},
volume={19},
number={1},
pages={21-27},
issn={1748-3395},
doi={10.1038/s41565-023-01491-3},
url={https://doi.org/10.1038/s41565-023-01491-3}
}

@article{Amico_2022,
  title = {Colloquium: Atomtronic circuits: From many-body physics to quantum technologies},
  author = {Amico, Luigi and Anderson, Dana and Boshier, Malcolm and Brantut, Jean-Philippe and Kwek, Leong-Chuan and Minguzzi, Anna and von Klitzing, Wolf},
  journal = {Rev. Mod. Phys.},
  volume = {94},
  issue = {4},
  pages = {041001},
  numpages = {30},
  year = {2022},
  month = {Nov},
  publisher = {American Physical Society},
  doi = {10.1103/RevModPhys.94.041001},
  url = {https://link.aps.org/doi/10.1103/RevModPhys.94.041001}
}

@Article{Chien_2015,
author={Chien, Chih-Chun
and Peotta, Sebastiano
and Di Ventra, Massimiliano},
title={Quantum transport in ultracold atoms},
journal={Nature Physics},
year={2015},
month={Dec},
day={01},
volume={11},
number={12},
pages={998-1004},
issn={1745-2481},
doi={10.1038/nphys3531},
url={https://doi.org/10.1038/nphys3531}
}

@article{Archak_2022,
  title = {Lyapunov equation in open quantum systems and non-Hermitian physics},
  author = {Purkayastha, Archak},
  journal = {Phys. Rev. A},
  volume = {105},
  issue = {6},
  pages = {062204},
  numpages = {19},
  year = {2022},
  month = {Jun},
  publisher = {American Physical Society},
  doi = {10.1103/PhysRevA.105.062204},
  url = {https://link.aps.org/doi/10.1103/PhysRevA.105.062204}
}

@Article{Sarkar_2025,
author={Sarkar, Saubhik
and Bayat, Abolfazl
and Bose, Sougato
and Ghosh, Roopayan},
title={Exponentially-enhanced quantum sensing with many-body phase transitions},
journal={Nature Communications},
year={2025},
month={Jun},
day={03},
volume={16},
number={1},
pages={5159},
issn={2041-1723},
doi={10.1038/s41467-025-60291-6},
url={https://doi.org/10.1038/s41467-025-60291-6}
}

@article{Raghunandan_2018,
  title = {High-Density Quantum Sensing with Dissipative First Order Transitions},
  author = {Raghunandan, Meghana and Wrachtrup, J\"org and Weimer, Hendrik},
  journal = {Phys. Rev. Lett.},
  volume = {120},
  issue = {15},
  pages = {150501},
  numpages = {6},
  year = {2018},
  month = {Apr},
  publisher = {American Physical Society},
  doi = {10.1103/PhysRevLett.120.150501},
  url = {https://link.aps.org/doi/10.1103/PhysRevLett.120.150501}
}

@article{Fern_2017,
  title = {Quantum sensing close to a dissipative phase transition: Symmetry breaking and criticality as metrological resources},
  author = {Fern\'andez-Lorenzo, Samuel and Porras, Diego},
  journal = {Phys. Rev. A},
  volume = {96},
  issue = {1},
  pages = {013817},
  numpages = {10},
  year = {2017},
  month = {Jul},
  publisher = {American Physical Society},
  doi = {10.1103/PhysRevA.96.013817},
  url = {https://link.aps.org/doi/10.1103/PhysRevA.96.013817}
}

@article{Zhou_2023,
  title = {Dynamical quantum phase transitions in a spinor Bose-Einstein condensate and criticality enhanced quantum sensing},
  author = {Zhou, Lu and Kong, Jia and Lan, Zhihao and Zhang, Weiping},
  journal = {Phys. Rev. Res.},
  volume = {5},
  issue = {1},
  pages = {013087},
  numpages = {10},
  year = {2023},
  month = {Feb},
  publisher = {American Physical Society},
  doi = {10.1103/PhysRevResearch.5.013087},
  url = {https://link.aps.org/doi/10.1103/PhysRevResearch.5.013087}
}

@book{Jishi_2013, place={Cambridge}, title={Feynman Diagram Techniques in Condensed Matter Physics}, publisher={Cambridge University Press}, author={Jishi, Radi A.}, year={2013}}

@book{Mattuck_2012,
  title={A guide to Feynman diagrams in the many-body problem},
  author={Mattuck, Richard D},
  year={2012},
  publisher={Courier Corporation}
}

@article{Prosen_2008,
  title = {Quantum Phase Transition in a Far-from-Equilibrium Steady State of an $XY$ Spin Chain},
  author = {Prosen, Toma\ifmmode \check{z}\else \v{z}\fi{} and Pi\ifmmode \check{z}\else \v{z}\fi{}orn, Iztok},
  journal = {Phys. Rev. Lett.},
  volume = {101},
  issue = {10},
  pages = {105701},
  numpages = {4},
  year = {2008},
  month = {Sep},
  publisher = {American Physical Society},
  doi = {10.1103/PhysRevLett.101.105701},
  url = {https://link.aps.org/doi/10.1103/PhysRevLett.101.105701}
}

@article{Prosen_2010,
doi = {10.1088/1367-2630/12/2/025016},
url = {https://doi.org/10.1088/1367-2630/12/2/025016},
year = {2010},
month = {feb},
publisher = {},
volume = {12},
number = {2},
pages = {025016},
author = {Prosen, Tomaž and Žunkovič, Bojan},
title = {Exact solution of Markovian master equations for quadratic Fermi systems: thermal baths, open XY spin chains and non-equilibrium phase transition},
journal = {New Journal of Physics}
}

@article{Bastidas_2012,
  title = {Nonequilibrium Quantum Phase Transitions in the Dicke Model},
  author = {Bastidas, V. M. and Emary, C. and Regler, B. and Brandes, T.},
  journal = {Phys. Rev. Lett.},
  volume = {108},
  issue = {4},
  pages = {043003},
  numpages = {5},
  year = {2012},
  month = {Jan},
  publisher = {American Physical Society},
  doi = {10.1103/PhysRevLett.108.043003},
  url = {https://link.aps.org/doi/10.1103/PhysRevLett.108.043003}
}

@article{Banchi_2014,
  title = {Quantum information-geometry of dissipative quantum phase transitions},
  author = {Banchi, Leonardo and Giorda, Paolo and Zanardi, Paolo},
  journal = {Phys. Rev. E},
  volume = {89},
  issue = {2},
  pages = {022102},
  numpages = {10},
  year = {2014},
  month = {Feb},
  publisher = {American Physical Society},
  doi = {10.1103/PhysRevE.89.022102},
  url = {https://link.aps.org/doi/10.1103/PhysRevE.89.022102}
}

@article{Dagvadorj_2015,
  title = {Nonequilibrium Phase Transition in a Two-Dimensional Driven Open Quantum System},
  author = {Dagvadorj, G. and Fellows, J. M. and Matyja\ifmmode \acute{s}\else \'{s}\fi{}kiewicz, S. and Marchetti, F. M. and Carusotto, I. and Szyma\ifmmode \acute{n}\else \'{n}\fi{}ska, M. H.},
  journal = {Phys. Rev. X},
  volume = {5},
  issue = {4},
  pages = {041028},
  numpages = {9},
  year = {2015},
  month = {Nov},
  publisher = {American Physical Society},
  doi = {10.1103/PhysRevX.5.041028},
  url = {https://link.aps.org/doi/10.1103/PhysRevX.5.041028}
}

@article{Carmichael_2015,
  title = {Breakdown of Photon Blockade: A Dissipative Quantum Phase Transition in Zero Dimensions},
  author = {Carmichael, H. J.},
  journal = {Phys. Rev. X},
  volume = {5},
  issue = {3},
  pages = {031028},
  numpages = {16},
  year = {2015},
  month = {Sep},
  publisher = {American Physical Society},
  doi = {10.1103/PhysRevX.5.031028},
  url = {https://link.aps.org/doi/10.1103/PhysRevX.5.031028}
}

@article{Archak_2017,
  title = {Nonequilibrium phase diagram of a one-dimensional quasiperiodic system with a single-particle mobility edge},
  author = {Purkayastha, Archak and Dhar, Abhishek and Kulkarni, Manas},
  journal = {Phys. Rev. B},
  volume = {96},
  issue = {18},
  pages = {180204},
  numpages = {5},
  year = {2017},
  month = {Nov},
  publisher = {American Physical Society},
  doi = {10.1103/PhysRevB.96.180204},
  url = {https://link.aps.org/doi/10.1103/PhysRevB.96.180204}
}

@article{Fitzpatrick_2017,
  title = {Observation of a Dissipative Phase Transition in a One-Dimensional Circuit QED Lattice},
  author = {Fitzpatrick, Mattias and Sundaresan, Neereja M. and Li, Andy C. Y. and Koch, Jens and Houck, Andrew A.},
  journal = {Phys. Rev. X},
  volume = {7},
  issue = {1},
  pages = {011016},
  numpages = {8},
  year = {2017},
  month = {Feb},
  publisher = {American Physical Society},
  doi = {10.1103/PhysRevX.7.011016},
  url = {https://link.aps.org/doi/10.1103/PhysRevX.7.011016}
}

@article{Hannukainen_2018,
  title = {Dissipation-driven quantum phase transitions and symmetry breaking},
  author = {Hannukainen, Julia and Larson, Jonas},
  journal = {Phys. Rev. A},
  volume = {98},
  issue = {4},
  pages = {042113},
  numpages = {11},
  year = {2018},
  month = {Oct},
  publisher = {American Physical Society},
  doi = {10.1103/PhysRevA.98.042113},
  url = {https://link.aps.org/doi/10.1103/PhysRevA.98.042113}
}

@article{Heugel_2019,
  title = {Quantum Transducer Using a Parametric Driven-Dissipative Phase Transition},
  author = {Heugel, Toni L. and Biondi, Matteo and Zilberberg, Oded and Chitra, R.},
  journal = {Phys. Rev. Lett.},
  volume = {123},
  issue = {17},
  pages = {173601},
  numpages = {6},
  year = {2019},
  month = {Oct},
  publisher = {American Physical Society},
  doi = {10.1103/PhysRevLett.123.173601},
  url = {https://link.aps.org/doi/10.1103/PhysRevLett.123.173601}
}

@article{Carollo_2020,
title = {Geometry of quantum phase transitions},
journal = {Physics Reports},
volume = {838},
pages = {1-72},
year = {2020},
note = {Geometry of quantum phase transitions},
issn = {0370-1573},
doi = {https://doi.org/10.1016/j.physrep.2019.11.002},
url = {https://www.sciencedirect.com/science/article/pii/S0370157319303655},
author = {Angelo Carollo and Davide Valenti and Bernardo Spagnolo}
}

@article{Zamora_2020,
  title = {Kibble-Zurek Mechanism in Driven Dissipative Systems Crossing a Nonequilibrium Phase Transition},
  author = {Zamora, A. and Dagvadorj, G. and Comaron, P. and Carusotto, I. and Proukakis, N. P. and Szyma\ifmmode \acute{n}\else \'{n}\fi{}ska, M. H.},
  journal = {Phys. Rev. Lett.},
  volume = {125},
  issue = {9},
  pages = {095301},
  numpages = {6},
  year = {2020},
  month = {Aug},
  publisher = {American Physical Society},
  doi = {10.1103/PhysRevLett.125.095301},
  url = {https://link.aps.org/doi/10.1103/PhysRevLett.125.095301}
}

@article{Minganti_2021,
doi = {10.1088/1367-2630/ac3db8},
url = {https://doi.org/10.1088/1367-2630/ac3db8},
year = {2021},
month = {dec},
publisher = {IOP Publishing},
volume = {23},
number = {12},
pages = {122001},
author = {Minganti, Fabrizio and Arkhipov, Ievgen I and Miranowicz, Adam and Nori, Franco},
title = {Continuous dissipative phase transitions with or without symmetry breaking},
journal = {New Journal of Physics}
}

@article{Archak_2021,
  title = {Subdiffusive Phases in Open Clean Long-Range Systems},
  author = {Purkayastha, Archak and Saha, Madhumita and Agarwalla, Bijay Kumar},
  journal = {Phys. Rev. Lett.},
  volume = {127},
  issue = {24},
  pages = {240601},
  numpages = {7},
  year = {2021},
  month = {Dec},
  publisher = {American Physical Society},
  doi = {10.1103/PhysRevLett.127.240601},
  url = {https://link.aps.org/doi/10.1103/PhysRevLett.127.240601}
}

@article{Gamayun_2021,
  title = {Nonequilibrium phase transition in transport through a driven quantum point contact},
  author = {Gamayun, Oleksandr and Slobodeniuk, Artur and Caux, Jean-S\'ebastien and Lychkovskiy, Oleg},
  journal = {Phys. Rev. B},
  volume = {103},
  issue = {4},
  pages = {L041405},
  numpages = {6},
  year = {2021},
  month = {Jan},
  publisher = {American Physical Society},
  doi = {10.1103/PhysRevB.103.L041405},
  url = {https://link.aps.org/doi/10.1103/PhysRevB.103.L041405}
}

@article{Saha_2023,
  title = {Universal Subdiffusive Behavior at Band Edges from Transfer Matrix Exceptional Points},
  author = {Saha, Madhumita and Agarwalla, Bijay Kumar and Kulkarni, Manas and Purkayastha, Archak},
  journal = {Phys. Rev. Lett.},
  volume = {130},
  issue = {18},
  pages = {187101},
  numpages = {6},
  year = {2023},
  month = {May},
  publisher = {American Physical Society},
  doi = {10.1103/PhysRevLett.130.187101},
  url = {https://link.aps.org/doi/10.1103/PhysRevLett.130.187101}
}

@Article{Kuzmin_2025,
author={Kuzmin, Roman
and Mehta, Nitish
and Grabon, Nicholas
and Mencia, Raymond A.
and Burshtein, Amir
and Goldstein, Moshe
and Manucharyan, Vladimir E.},
title={Observation of the Schmid--Bulgadaev dissipative quantum phase transition},
journal={Nature Physics},
year={2025},
month={Jan},
day={01},
volume={21},
number={1},
pages={132-136},
doi={10.1038/s41567-024-02695-7},
url={https://doi.org/10.1038/s41567-024-02695-7}
}

@article{Matsumoto_2025,
  title = {Dissipative quantum phase transitions monitored by current fluctuations},
  author = {Matsumoto, Masataka and Cai, Zi and Baggioli, Matteo},
  journal = {Phys. Rev. A},
  volume = {112},
  issue = {1},
  pages = {012226},
  numpages = {12},
  year = {2025},
  month = {Jul},
  publisher = {American Physical Society},
  doi = {10.1103/2fw8-bsjy},
  url = {https://link.aps.org/doi/10.1103/2fw8-bsjy}
}

@article{Dhar_2006_1,
  title = {Nonequilibrium Green's function formalism and the problem of bound states},
  author = {Dhar, Abhishek and Sen, Diptiman},
  journal = {Phys. Rev. B},
  volume = {73},
  issue = {8},
  pages = {085119},
  numpages = {14},
  year = {2006},
  month = {Feb},
  publisher = {American Physical Society},
  doi = {10.1103/PhysRevB.73.085119},
  url = {https://link.aps.org/doi/10.1103/PhysRevB.73.085119}
}

@Article{Dhar_2006_2,
author={Dhar, Abhishek
and Roy, Dibyendu},
title={Heat Transport in Harmonic Lattices},
journal={Journal of Statistical Physics},
year={2006},
month={Nov},
day={01},
volume={125},
number={4},
pages={801-820},
issn={1572-9613},
doi={10.1007/s10955-006-9235-3},
url={https://doi.org/10.1007/s10955-006-9235-3}
}

\end{document}